\documentclass[%
 reprint,
 amsmath,amssymb,
 aps,
pra,
]{revtex4-1}

\usepackage{subfigure}
\usepackage{graphicx}
\usepackage{dcolumn}
\usepackage{bm}
\usepackage{amsmath,amsfonts,amssymb}

\begin{document}


\title{Efficient Generation of the Triplet Bell state Between Coupled Spins Using Transitionless Quantum Driving and Optimal Control}

\author{Dionisis Stefanatos}
\email{dionisis@post.harvard.edu}
\author{Emmanuel Paspalakis}%
\affiliation{Materials Science Department, School of Natural Sciences, University of Patras, Patras 26504, Greece}

\date{\today}

\begin{abstract}

We consider a pair of coupled spins with Ising interaction in z-direction and study the problem of generating efficiently the triplet Bell state. We initially analyze the transitionless quantum driving shortcut to adiabaticity method and point out its limitations when the available duration approaches zero. In this short time limit we explicitly calculate the fidelity of the method and find it to be much lower than unity, no matter how large become the available control fields. We find that there is a lower bound on the necessary time to complete this transfer, set by the finite value of the interaction between the spins. We then use numerical optimal control to find bang-bang pulse sequences, as well as, smooth controls, which can generate high levels of the target Bell state in the minimum possible time. The results of the present work are not restricted only to spin systems, but is expected to find also applications in other physical systems which can be modeled as interacting spins, such as, for example, coupled quantum dots.

\end{abstract}

\maketitle

\section{Introduction}

\label{sec:intro}


A prototype quantum system which plays a central role in quantum information processing consists of two spins interacting through Ising coupling in $z$-direction. This system has been used in the early demonstrations of quantum algorithms using Nuclear Megnetic Resonance (NMR) experiments \cite{Chuang98,Jones98} and can serve as a building block for quantum computation.

An important problem related to this system is the efficient creation of the triplet Bell state $\frac{1}{\sqrt{2}}(|\downarrow\rangle_1|\uparrow\rangle_2+|\uparrow\rangle_1|\downarrow\rangle_2)$ when starting from the spin-down state $|\downarrow\rangle_1|\downarrow\rangle_2$ \cite{Unanyan01}, which has been recently attracted considerable attention \cite{Paul16,Zhang17,Yu18}. This transfer is interesting not only theoretically but also for practical purposes. For example, in the case of two coupled quantum dots, a system described by a similar Hamiltonian with two coupled spins \cite{Kis04}, this transfer corresponds to the creation of the single-exciton symmetric state when starting from the vacuum state \cite{Kis04,Quiroga99a}. For the efficient generation of the triplet Bell state in such a pair of spins, a technique based on rapid adiabatic passage \cite{Vitanov01} has been proposed \cite{Unanyan01}, according to which a suitably chosen time-dependent external field drives the pair adiabatically from the spin-down state to the target Bell state. The advantage of the adiabatic passage method is its robustness against system imperfections, for example field inhomogeneities. Its inherent drawback is the long necessary time to complete the transfer, which becomes particularly important in the presence of dissipation.

In order to speed up adiabatic quantum dynamics, a series of closely related methods have been proposed over the past few years \cite{Demirplak03,Berry09,Motzoi09,Chen10a,Masuda10,Deffner14}. These techniques are collectively referred as \emph{Shortcuts to Adiabaticity}. The main idea behind them is that the system arrives at the same final state as with a slow adiabatic process, but it doesn't necessarily follow the instantaneous adiabatic eigenstates at intermediate times. These methods have been exploited to accelerate quantum adiabatic evolution in
a wide range of applications. These include the fast cooling and transport of atoms \cite{Chen10a,Chen11a,Ness18}, Bose-Einstein condensates \cite{Schaff11} and trapped ions \cite{An16}, the efficient manipulation of two-, three- and four-level quantum systems \cite{Chen10,Chen11,Gungordu12,Li18}, the effective generation of entanglement between ultracold gases \cite{Stefanatos18a,Fogarty18} and exciton-polaritons \cite{Stefanatos18b}, the design of waveguides and photonic lattices \cite{Tseng12,Stefanatos14a}, the optimization of quantum heat engines \cite{Deng13,delCampo14,Zheng16,Kosloff17,Deng17}, the fast optomechanical cooling \cite{Li11} and quantum computation \cite{Santos15,Palmero17}, and even the control of mechanical systems \cite{Gonzalez17}. Several studies have also been devoted to the control of spin dynamics \cite{Takahashi13,Paul16,Paul16,Setiawan17,Shi17,Shi17,Zhang17,Yu18}. For the efficient generation of the triplet Bell state in a pair of Ising coupled spins, two shortcut methods have been used. The first is \emph{Transitionless Quantum Driving} (TQD) \cite{Paul16}, where an extra term is added to the Hamiltonian so the system follows the instantaneous eigenstates of the original Hamiltonian \cite{Demirplak03,Berry09}, and the second is \emph{Lewis-Riesenfeld Invariant} (LRI) inverse engineering \cite{Paul16,Zhang17,Yu18}, where the system evolves along the eigenstates of a motion invariant \cite{Lewis69,Chen10}.

In the present article, we initially re-examine the TQD method and identify its limitations. Specifically, we show that when the available duration approaches zero, the fidelity of the target Bell state approaches the constant value $\sin^2{(\pi/\sqrt{2})}/2\approx 0.3166$. Consequently, the fidelity of this transfer cannot achieve values close to unity in arbitrarily short times, as claimed in \cite{Paul16}, despite the fact that one of the controls becomes actually a delta pulse in the short time limit. The short time behavior that we derive here for the TQD shortcut is analogous to that of the LRI shortcut obtained in \cite{Zhang17}. There is a lower bound on the necessary time to complete this transfer, set by the finite value of the interaction between the spins. Having determined the limits of the TQD method, we next use numerical optimal control to find bang-bang pulse sequences, as well as, smooth controls which can generate high levels of the desired Bell state in the minimum possible time. The current work follows a series of analytical and numerical studies on the optimal control of spin dynamics \cite{Khaneja01,Stefanatos04,Stefanatos05,Boscain06,Stefanatos09,Lapert10,Burgarth10,Wang10,Moore12,Hegerfeldt13,Mukherjee13,Dolde14,Watts15,Aiello15,Goodwin16,Goodwin16,VanDamme18}. Yet, the results of the present research is expected to find also applications in other physical systems which can be modeled as interacting spins, for example, coupled quantum dots \cite{Quiroga99a,Lovett03a,Kis04,Aty07a,Creatore12}, as the resulted Hamiltonian can also occur in these systems as well.

The structure of the paper is as follows. In the next section we present the model of two Ising interacting spins and briefly discuss the rapid adiabatic passage method for generating the triplet Bell state. In Section \ref{sec:TQD} we consider the TQD shortcut and point out its limitations, while in Section \ref{sec:OC} we use numerical optimal control to create sufficient levels of the target state within short times. Section \ref{sec:conclusion} concludes this work.

\section{A spin pair with Ising coupling in a time dependent magnetic field}

\label{sec:model}

We consider a pair of spin-$\frac{1}{2}$ particles with an Ising interaction along the $z$-axis, which is embedded in a time dependent magnetic field $\mathbf{B}(t)=[B_x(t), B_y(t), B_z(t)]$. The corresponding Hamiltonian is \cite{Unanyan01}
\begin{equation}
\label{Hamiltonian}
\hat{H}(t)=4\xi \hat{S}_{1z}\hat{S}_{2z}+\mu\mathbf{B}(t)\cdot(\hat{\mathbf{S}}_1+\hat{\mathbf{S}}_2),
\end{equation}
where $\xi>0$ denotes the strength of the Ising coupling, $\mu$ is the gyromagnetic ratio, and $\hat{\mathbf{S}}_i=(S_{ix},S_{iy},S_{iz})$ is the spin operator for the $i$th particle, $i=1,2$, with elements proportional to the Pauli matrices.

A suitable orthonormal basis consists of the triplet
\begin{subequations}
\label{triplet}
\begin{eqnarray}
|\psi_{\downdownarrows}\rangle &=& |\downarrow\rangle_1|\downarrow\rangle_2,\label{down}\\
|\psi^+_{\downarrow\uparrow}\rangle &=&\frac{1}{\sqrt{2}}(|\downarrow\rangle_1|\uparrow\rangle_2+|\uparrow\rangle_1|\downarrow\rangle_2),\label{entangled}\\
|\psi_{\upuparrows}\rangle &=&|\uparrow\rangle_1|\uparrow\rangle_2 , \label{up}
\end{eqnarray}
\end{subequations}
and the singlet
\begin{equation}
\label{singlet}
|\psi^-_{\downarrow\uparrow}\rangle=\frac{1}{\sqrt{2}}(|\downarrow\rangle_1|\uparrow\rangle_2-|\uparrow\rangle_1|\downarrow\rangle_2),
\end{equation}
states, where $|\uparrow\rangle, |\downarrow\rangle$ denote the spin-up and spin-down states respectively. It can be easily verified that the singlet state, with total spin 0, is decoupled from the triplet states, characterized by total spin 1. Within the triplet manifold, Hamiltonian (\ref{Hamiltonian}) can be expressed in matrix form as \cite{Unanyan01}
\begin{equation}
\label{Ha}
H_a(t)=
\left[
\begin{array}{ccc}
\xi-\beta_z & \frac{1}{\sqrt{2}}(\beta_x+i\beta_y)  & 0\\
\frac{1}{\sqrt{2}}(\beta_x-i\beta_y) & -\xi & \frac{1}{\sqrt{2}}(\beta_x+i\beta_y) \\
0 & \frac{1}{\sqrt{2}}(\beta_x-i\beta_y)  & \xi+\beta_z
\end{array}
\right],
\end{equation}
where $\mathbf{\beta}=\mu\mathbf{B}$. Observe that the triplet states are coupled through the transverse $xy$-magnetic field. The corresponding probability amplitudes $\mathbf{a}(t)=[a_1(t), a_2(t), a_3(t)]^T$ obey the Schr\"{o}dinger equation ($\hbar=1$)
\begin{equation}
\label{SchrodingerA}
i\frac{d}{dt}\mathbf{a}(t)=H_a(t)\mathbf{a}(t).
\end{equation}
We consider that initially the system is in the unentangled spin-down state $|\psi_{\downdownarrows}\rangle$ and our goal is to find the appropriate magnetic field which drives it efficiently to the maximally entangled Bell state $|\psi^+_{\downarrow\uparrow}\rangle$.

Following \cite{Unanyan01} we choose a rotating transverse magnetic field
\begin{subequations}
\label{operators}
\begin{eqnarray}
\beta_x(t)&=&\Omega(t)\cos{\omega t},\label{Bx}\\
\beta_y(t)&=&\Omega(t)\sin{\omega t}.\label{By}
\end{eqnarray}
\end{subequations}
Under this field, the transformed probability amplitudes
$c_1(t)=a_1(t)e^{-i(\omega+\xi)t}, c_2(t)=a_2(t)e^{-i\xi t}, c_3(t)=a_3(t)e^{i(\omega-\xi)t}$
obey the equation
\begin{equation}
\label{SchrodingerC}
i\frac{d}{dt}\mathbf{c}(t)=H_c(t)\mathbf{c}(t),
\end{equation}
where
\begin{equation}
\label{Hc}
H_c(t)=
\left[
\begin{array}{ccc}
\Delta(t) & \frac{1}{\sqrt{2}}\Omega(t)  & 0\\
\frac{1}{\sqrt{2}}\Omega(t) & 0 & \frac{1}{\sqrt{2}}\Omega(t) \\
0 & \frac{1}{\sqrt{2}}\Omega(t)  & 4\xi-\Delta(t)
\end{array}
\right]
\end{equation}
and the detuning $\Delta(t)$ is defined as \cite{Unanyan01}
\begin{equation}
\label{detuning}
\Delta(t)=2\xi+\omega-\beta_z(t).
\end{equation}

Observe from (\ref{Hc}) that the transverse field $\Omega(t)$ couples both $|\psi_{\downdownarrows}\rangle, |\psi^+_{\downarrow\uparrow}\rangle$ and  $|\psi^+_{\downarrow\uparrow}\rangle, |\psi_{\upuparrows}\rangle$. In order to efficiently achieve the desired transfer $|\psi_{\downdownarrows}\rangle\rightarrow|\psi^+_{\downarrow\uparrow}\rangle$, while simultaneously suppressing the undesirable transfer $|\psi^+_{\downarrow\uparrow}\rangle\rightarrow|\psi_{\upuparrows}\rangle$, the authors of \cite{Unanyan01} employed an adiabatic rapid passage technique. They used $\beta_z(t)=At$, i.e a linear variation of the detuning, and a gaussian $\Omega(t)$ centered at the point where $|\psi_{\downdownarrows}\rangle, |\psi^+_{\downarrow\uparrow}\rangle$ become degenerate, while $|\psi_{\upuparrows}\rangle$ is far detuned. With this adiabatic method the desired transfer is accomplished in a robust way, but it requires a sufficient amount of time, which might be a drawback in the presence of dissipation. In the following sections we use two methods to reduce the necessary transfer time, namely TQD and optimal control.

\section{Transitionless quantum driving}

\label{sec:TQD}

Following \cite{Unanyan01,Paul16,Zhang17,Yu18}, we derive the TQD shortcut to adiabaticity for the two-level system describing the interaction between the states $|\psi_{\downdownarrows}\rangle, |\psi^+_{\downarrow\uparrow}\rangle$ and than test it for the full three-level system described by Hamiltonian (\ref{Hc}). From Eq. (\ref{Hc}) and after a simple unitary transformation we obtain the following Hamiltonian for the two-level interaction
\begin{eqnarray}
H_0(t)&=&
\frac{1}{2}
\left[
\begin{array}{cc}
\Delta(t) & \sqrt{2}\Omega(t)\\
\sqrt{2}\Omega(t) & -\Delta(t)
\end{array}
\right]
\Rightarrow\nonumber\\
\hat{H}_0(t)&=&\Delta(t)\hat{S}_z+\sqrt{2}\Omega(t)\hat{S}_x,\label{H0}
\end{eqnarray}
where $\hat{S}_x, \hat{S}_z$ are proportional to the Pauli spin matrices. The goal is to transfer the population from the initial to the final state following the adiabatic paths of Hamiltonian (\ref{H0}). But the eigenstates of this Hamiltonian are time-dependent, thus a transformation to the adiabatic basis leads to non-diagonal diabatic terms which can be neglected only in the adiabatic (long time) limit, and this is the case where the system follows the instantaneous eigenstates. The idea behind TQD is to add an extra term $\hat{H}_{cd}(t)$ to the Hamiltonian to cancel the diabatic effects, so the system can follow the instantaneous eigenstates of the reference Hamiltonian $\hat{H}_0(t)$ even for arbitrarily short times. In order to find the counterdiabatic term, we find first the instantaneous eigenvalues and eigenstates of the two-level Hamiltonian (\ref{H0}).

If we parametrize $\Delta,\Omega$ as
\begin{subequations}
\label{parametrized_controls}
\begin{eqnarray}
\Delta(t)&=&E_0\cos{\theta},\\
\Omega(t)&=&\frac{E_0}{\sqrt{2}}\sin{\theta},
\end{eqnarray}
\end{subequations}
with time dependent $E_0(t), \theta(t)$, then
\begin{equation}\label{parametrized_H0}
H_0=\frac{E_0}{2}
\left(\begin{array}{cc}
    \cos\theta & \sin\theta \\ \sin\theta & -\cos\theta
  \end{array}\right),
\end{equation}
with instantaneous eigenvalues
\begin{equation}\label{eigenvalus}
E_{\pm}=\pm\frac{E_0}{2} \, ,
\end{equation}
and normalized eigenvectors
\begin{subequations}
\label{eigenvectors}
\begin{eqnarray}
|\phi_{+}(t)\rangle&=&
\left(\begin{array}{c}
    \cos{\frac{\theta}{2}}\\
    \sin{\frac{\theta}{2}}
\end{array}\right),\label{plus}\\
|\phi_{-}(t)\rangle&=&
\left(\begin{array}{c}
    \sin\frac{\theta}{2}\\
    -\cos{\frac{\theta}{2}}
\end{array}\right).\label{minus}
\end{eqnarray}
\end{subequations}
If $\hat{H}_0(t)$ is varied slowly, then the system follows the approximate adiabatic solutions
\begin{equation}\label{reference_solution}
|\psi_\pm^0(t)\rangle=e^{i\xi_{\pm}(t)}|\phi_\pm(t)\rangle,
\end{equation}
where the phases are
\begin{eqnarray}
\xi_\pm(t)&=&-\int_0^tdt'E_{\pm}(t')+i\int_0^tdt'\langle\phi_\pm(t')|\dot{\phi}_\pm(t')\rangle\nonumber\\
          &=&-\int_0^tdt'E_{\pm}(t'),\label{xi}
\end{eqnarray}
since the inner product term in Eq. (\ref{xi}) is zero.

The counterdiabatic Hamiltonian is given by \cite{Demirplak03,Berry09}
\begin{eqnarray}
\label{counter_diabatic}
\hat{H}_{cd}(t)&=&i\sum_{n=\pm}\Big[|\dot{\phi}_n(t)\rangle\langle\phi_n(t)|\nonumber\\
         & &-\langle\phi_n(t)|\dot{\phi}_n(t)\rangle|\phi_n(t)\rangle\langle\phi_n(t)|\Big]\nonumber\\
         &=&i\sum_{n=\pm}|\dot{\phi}_n(t)\rangle\langle\phi_n(t)|\nonumber\\
         &=&\dot{\theta}S_y,
\end{eqnarray}
since the term in the second line of Eq. (\ref{counter_diabatic}) is zero. Under the total Hamiltonian
\begin{equation}\label{tracking_Hamiltonian}
\hat{H}(t)=\hat{H}_0(t)+\hat{H}_{cd}(t),
\end{equation}
the state $|\psi\rangle$ of the system, which satisfies the Schr\"{o}dinger equation
\begin{equation}\label{counterdiabatic}
  i\frac{\partial}{\partial t}|\psi(t)\rangle=\hat{H}(t)|\psi(t)\rangle,
\end{equation}
follows exactly the adiabatic solutions (\ref{reference_solution}) of the reference Hamiltonian $\hat{H}_0(t)$, no matter how short is the duration $T$ of the evolution.

The introduction of the extra term $\hat{H}_{cd}=\dot{\theta}S_y$ in system's Hamiltonian is in generally undesirable. However, there is an alternative method to implement the shortcut with a Hamiltonian of the same form as $\hat{H}_0$ \cite{Ibanez12,Martinez14}.
Consider the unitary transformation
\begin{equation}
\label{psiI}
|\psi'(t)\rangle=\hat{U}^\dagger(t)|\psi(t)\rangle
\end{equation}
with
\begin{equation}\label{B}
U(t)=e^{-ib(t)\hat{S}_z},
\end{equation}
where $b(t)$ is a real function of time to be determined. The transformed state
obeys the alternative dynamics
\begin{equation}\label{SchrodingerI}
  i\frac{\partial}{\partial t}|\psi'(t)\rangle=\hat{H}'(t)|\psi'(t)\rangle,
\end{equation}
with the modified Hamiltonian
\begin{eqnarray}
\hat{H}'(t)&=& \hat{U}^\dagger(t)\hat{H}(t)\hat{U}(t)-i\hat{U}^\dagger(t)\frac{d}{dt}\hat{U}(t)\nonumber\\
           &=& (E_0\cos\theta-\dot{b})\hat{S}_z+(E_0\sin\theta\cos b+\dot{\theta}\sin b)\hat{S}_x\nonumber\\
           &+&(\dot{\theta}\cos b-E_0\sin\theta\sin b)\hat{S}_y.\label{HIB}
\end{eqnarray}
The choice
\begin{equation}\label{b}
\tan b =\frac{\dot{\theta}}{E_0\sin\theta}
\end{equation}
eliminates the undesirable extra term proportional to $\hat{S}_y$ in Eq. (\ref{HIB}) and we finally get
\begin{equation}\label{HItwolevel}
\hat{H}'(t) = \Delta'(t)\hat{S}_z+\sqrt{2}\Omega'(t)\hat{S}_x,
\end{equation}
which has the same form as Eq. (\ref{H0}) but with modified controls \cite{Martinez14}
\begin{subequations}
\label{mod_controls}
\begin{eqnarray}
&\Delta'(t)&=E_0\cos\theta-\dot{b}\label{UI}\\ &=&\frac{E_0^3\sin^2\theta\cos\theta+\dot{E}_0\dot{\theta}\sin\theta+E_0(2\dot{\theta}^2\cos\theta-\ddot{\theta}\sin\theta)}{E_0^2\sin^2\theta+\dot{\theta}^2},\nonumber\\
&\Omega'(t)& = \frac{E_0\sin\theta\cos b+\dot{\theta}\sin b}{\sqrt{2}} = \sqrt{\frac{E_0^2\sin^2\theta+\dot{\theta}^2}{2}}.\label{JI}
\end{eqnarray}
\end{subequations}

\begin{figure}[t]
 \centering
		\begin{tabular}{cc}
     	\subfigure[$\ $]{
	            \label{fig:phi}
	            \includegraphics[width=.5\linewidth]{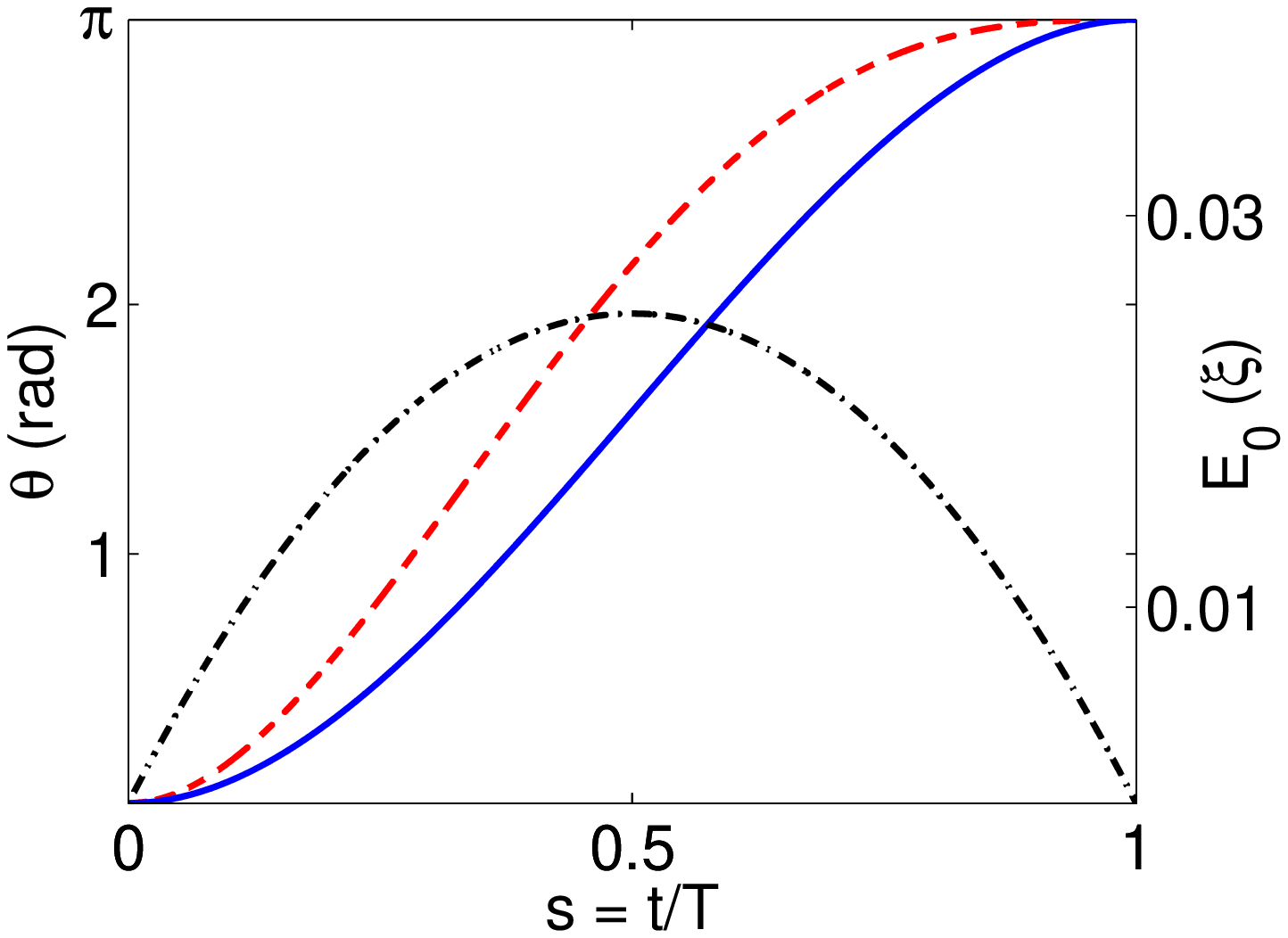}} &
       \subfigure[$\ $]{
	            \label{fig:shortcut_fidelity}
	            \includegraphics[width=.5\linewidth]{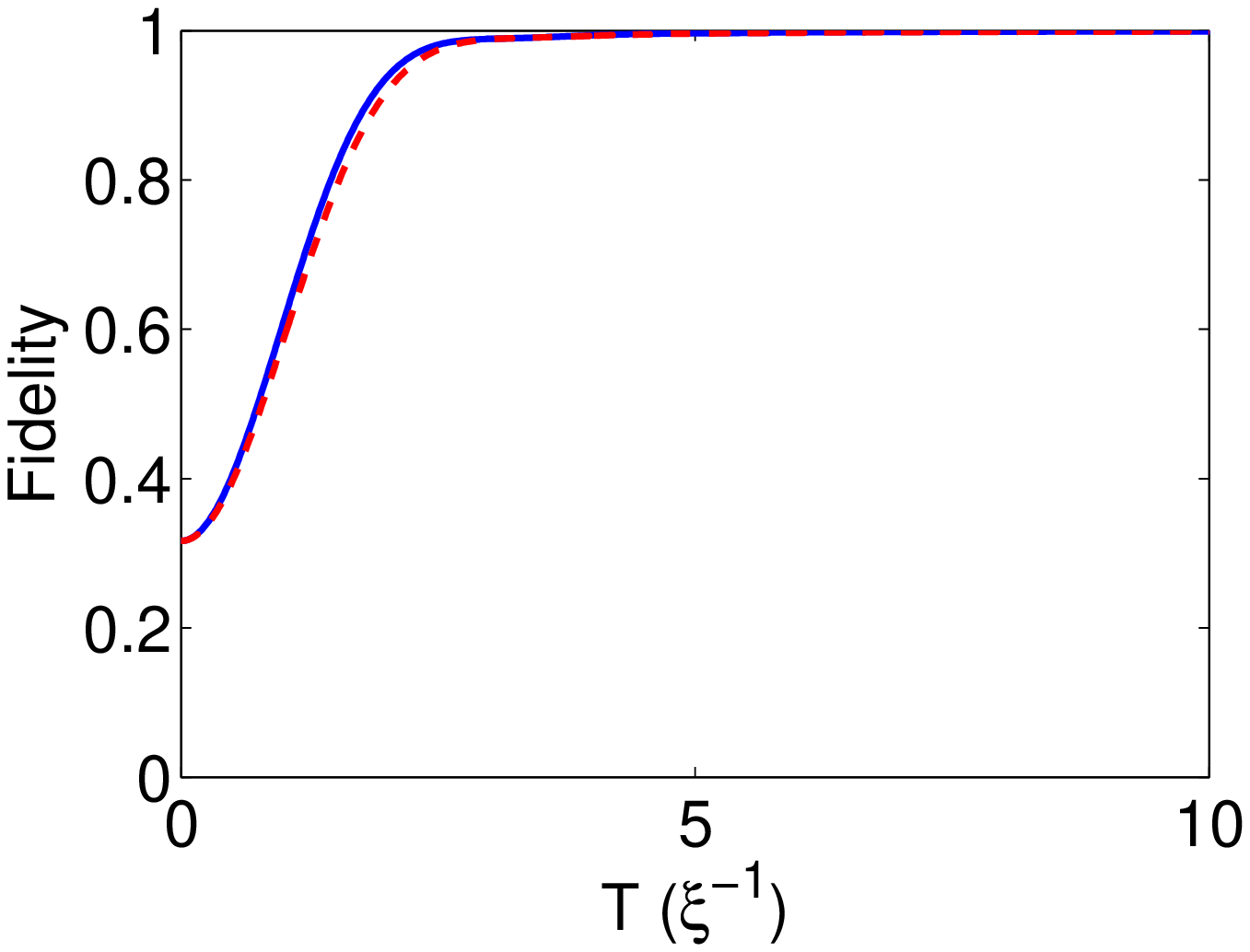}} \\
	    \subfigure[$\ $]{
	            \label{fig:delta_shortcut}
	            \includegraphics[width=.5\linewidth]{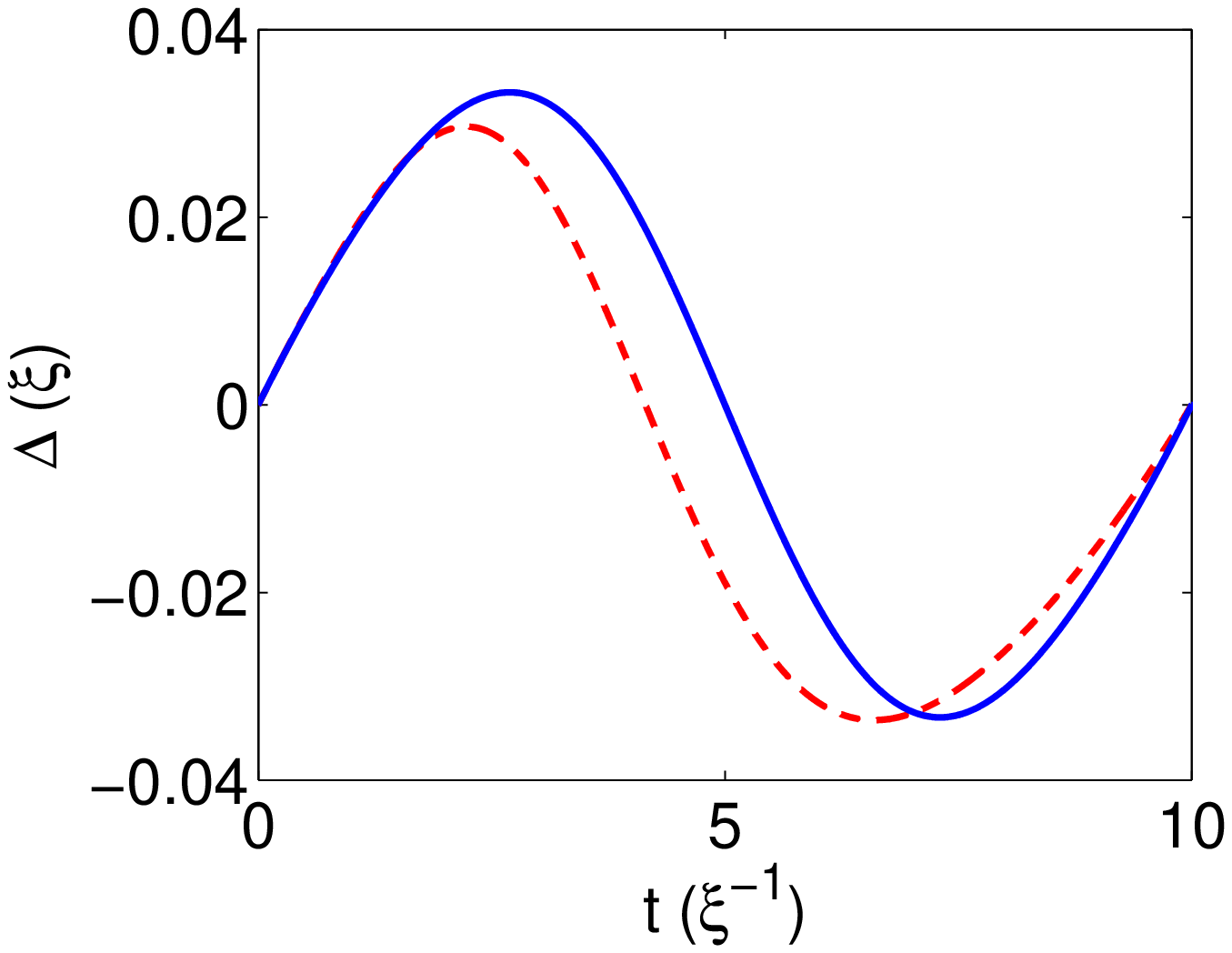}} &
       \subfigure[$\ $]{
	            \label{fig:omega_shortcut}
	            \includegraphics[width=.5\linewidth]{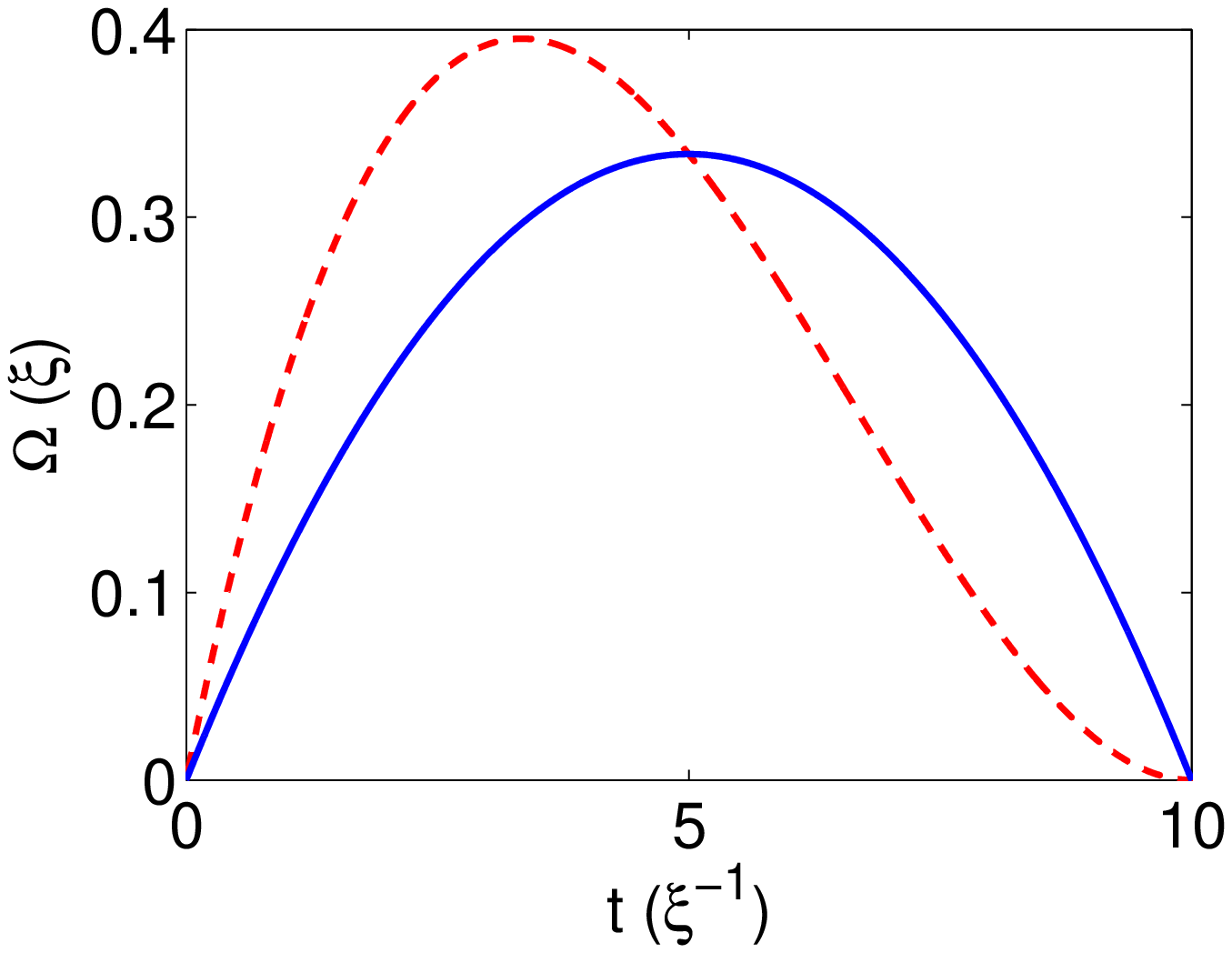}} \\
       \subfigure[$\ $]{
	            \label{fig:end}
	            \includegraphics[width=.5\linewidth]{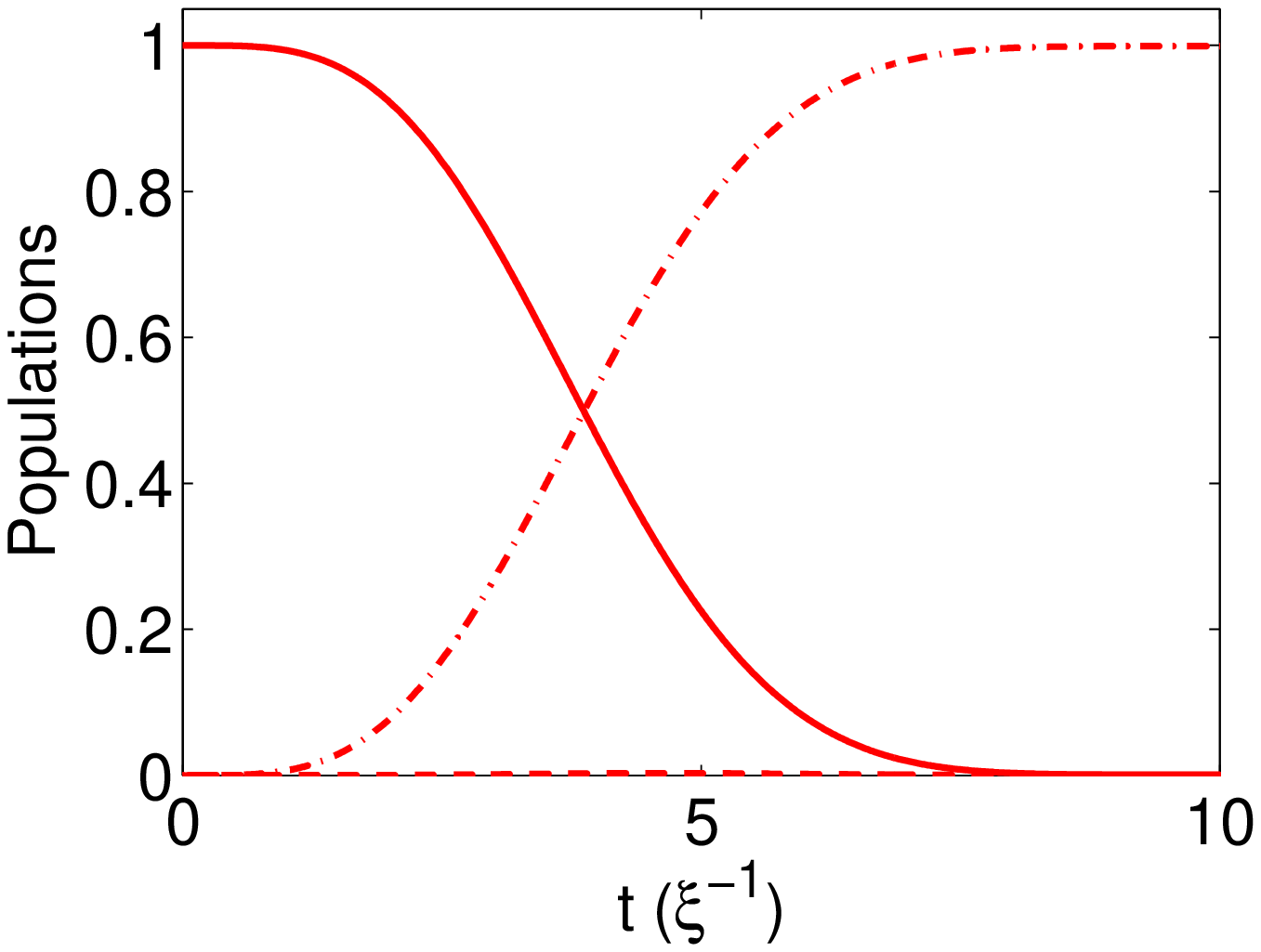}} &
       \subfigure[$\ $]{
	            \label{fig:middle}
	            \includegraphics[width=.5\linewidth]{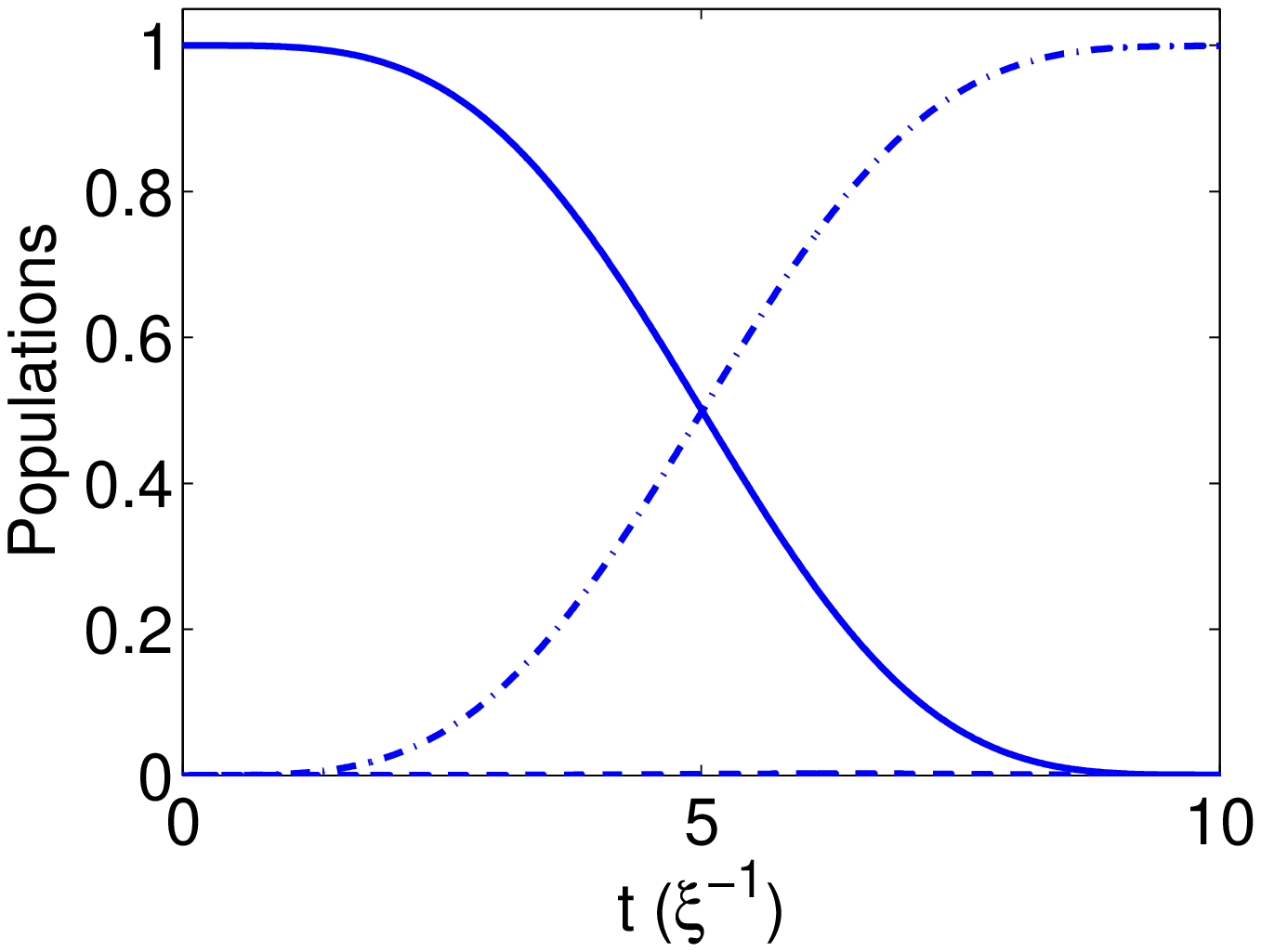}}
		\end{tabular}
\caption{(Color online) Red dashed line corresponds to the nonsymmetric shortcut (\ref{phins}), blue solid line to the symmetric one (\ref{phis}). (a) Reference angle $\theta$ (left vertical axis) for the two shortcuts and common $E_0$ (right vertical axis, black dashed-dotted line), as functions of normalized time $s=t/T$. (b) Fidelity as a function of duration $T$ for the two shortcuts. In the limit $T\rightarrow 0$ the fidelity approaches the value given in Eq. (\ref{eta}). (c) Detuning $\Delta(t)$ for duration $T=10\xi^{-1}$. (d) Rabi frequency $\Omega(t)$ for $T=10\xi^{-1}$. (e) Time evolution of populations for the nonsymmetric shortcut when $T=10\xi^{-1}$. The final fidelity is $0.9991$. (f) Time evolution of populations for the symmetric shortcut when $T=10\xi^{-1}$. The final fidelity is $0.9993$.}
\label{fig:shortcuts}
\end{figure}

We first derive the shortcut for the modified dynamics (\ref{SchrodingerI}) of the transformed state $|\psi'(t)\rangle$ and then explain why it is also a shortcut for the original dynamics (\ref{counterdiabatic}) of state $|\psi(t)\rangle$.
We find the appropriate functions of time $\theta(t), E_0(t)$ which determine the reference adiabatic path. In order to satisfy the initial and final conditions of the transfer, from Eq. (\ref{eigenvectors}) it is evident that the evolution should take place along the adiabatic solution $|\psi_+^0(t)\rangle$ with boundary conditions for $\theta$
\begin{subequations}
\label{boundary_phi}
\begin{eqnarray}
\theta(0)&=&0,\\
\theta(T)&=&\pi.
\end{eqnarray}
\end{subequations}
The smoothness conditions
\begin{equation}\label{boundary_dotphi}
\dot{\theta}(0)=\dot{\theta}(T)=0
\end{equation}
also imply that $H_{cd}(0)=H_{cd}(T)=0$, i.e. the extra term in the counterdiabatic Hamiltonian (\ref{counter_diabatic}) vanishes at the boundary times.
In Ref. \cite{Martinez14} the following extra condition
\begin{equation}\label{boundary_ddotphi}
\ddot{\theta}(T)=0
\end{equation}
is used, which is actually not necessary for the population inversion that we want to accomplish here.
Using a polynomial to interpolate the function $\theta(t)$ at intermediate times and imposing on it the above boundary conditions, we find ($s=t/T$)
\begin{subequations}
\label{phi}
\begin{eqnarray}
\label{phis}\theta_s(s)&=&\pi s^2(3-2s),\\
\label{phins}\theta_{ns}(s)&=&\pi s^2(3s^2-8s+6),
\end{eqnarray}
\end{subequations}
where $\theta_{s}$ satisfies only the symmetric boundary conditions (\ref{boundary_phi}), (\ref{boundary_dotphi}) which lead to $\theta_{s}(T/2)=\pi/2$, while $\theta_{ns}$ additionally incorporates the nonsymmetric condition (\ref{boundary_ddotphi}).
In order to make the control fields vanish at the boundary times, we impose the conditions
\begin{equation}\label{boundary_E}
E_0(0)=E_0(T)=0.
\end{equation}
The simple polynomial
\begin{equation}
\label{E}
E_0(s)=es(1-s),
\end{equation}
where $e$ is some constant, satisfies the above conditions and also assures that $E_0(T/2)\neq 0$. In Fig. \ref{fig:phi} we plot both $\theta_s(s)$ (blue solid line) and $\theta_{ns}(s)$ (red dashed line), as well as $E_0(s)$ (black dashed-dotted line) with $e=0.1\xi$.

We next show that the shortcut derived above inverts the populations also in the original picture described by state $|\psi(t)\rangle$. We find the unitary transformation connecting the states $|\psi(t)\rangle, |\psi'(t)\rangle$ at the boundary times $t=0,T$.
Using Eqs. (\ref{phi}), (\ref{E}) in (\ref{b}), we can evaluate the boundary values of $b(t)$ in the limits $s\rightarrow 0, 1$,
\begin{equation}\label{boundary_b}
b(0)=b(T)=\frac{\pi}{2}.
\end{equation}
From Eqs. (\ref{B}) and (\ref{boundary_b}) we obtain
\begin{equation}\label{boundary_B}
U(0)=U(T)=\left(\begin{array}{cc}e^{-i\pi/4} & 0 \\ 0 & e^{i\pi/4}\end{array}\right),
\end{equation}
thus $|\psi'(0)\rangle=e^{i\pi/4}|\psi(0)\rangle$ and $|\psi'(T)\rangle=e^{-i\pi/4}|\psi(T)\rangle$, and obviously the counterdiabatic shortcut inverts the population also in the original picture. Working analogously we can also find
\begin{equation}\label{boundary_dotb}
\dot{b}(0)=\dot{b}(T)=0
\end{equation}
which, along with the boundary conditions for $\theta,\dot{\theta}$, imply that $\hat{H}'(t_b)=\hat{H}(t_b)=\hat{H}_0(t_b)$, where $t_b=0, T$. The bottom line of the above analysis is that, if we apply the modified controls (\ref{mod_controls}) in the two-level Hamiltonian $\hat{H}_0$ (\ref{H0}), then the desired population inversion is accomplished along the adiabatic path $|\psi_+^0(t)\rangle$ (\ref{reference_solution}). This transfer can in principle be completed in arbitrarily short times $T$. In the rest of this section we drop the prime from the left hand side of Eq. (\ref{mod_controls}) and use the symbols $\Delta(t),\Omega(t)$ to denote the modified controls.

We now move to evaluate the performance of the method when applied to the three-level system (\ref{SchrodingerC}) described by Hamiltonian (\ref{Hc}). In Fig. \ref{fig:shortcut_fidelity} we plot the fidelity $|c_2(T)|^2$ as a function of duration $T$, for both the symmetric (blue solid line) and the nonsymmetric (red dashed line) TQD shortcuts given in Eq. (\ref{phi}), when the corresponding controls (\ref{mod_controls}) are applied in the three-level Hamiltonian $H_c$. Observe that, as $T\rightarrow 0$, the fidelity approaches a value much less than 1. A similar behavior has been observed for shortcuts designed using LRIs, see Fig. 7 in Ref. \cite{Zhang17}, and it is attributed to the finite value of the coupling $\xi$. We can actually calculate explicitly the short time fidelity limit. Note that the time derivative of a function $f(t)$ can be expressed as $\dot{f}=df/dt=(1/T)df/ds=f'/T$, where $f'=df/ds$ denotes the derivative with respect to normalized time $s=t/T$. Using this recipe, it is not hard to show that in the short time limit $T\rightarrow 0$ the controls (\ref{mod_controls}) become
\begin{subequations}
\label{limit_controls}
\begin{eqnarray}
\Delta(t)&=&\frac{E'_0\theta'\sin\theta+E_0(2\theta'^2\cos\theta-\theta''\sin\theta)}{\theta'^2},\label{limit_D}\\
\Omega(t)&=&\frac{1}{T\sqrt{2}}\frac{d\theta}{ds}.\label{limit_W}
\end{eqnarray}
\end{subequations}
Observe that $\Omega(t)$ becomes a delta pulse as $T\rightarrow 0$, while $\Delta(t)$ remains finite. In this limit thus we can keep only the terms proportional to $\Omega$ in Hamiltonian $H_c$ (\ref{Hc}). Under this approximation, we integrate Schr\"{o}dinger equation (\ref{SchrodingerC}) starting from $\mathbf{c}(0)=(1, 0, 0)^T$ and find
\begin{equation*}
c_2(T)=\frac{-i}{\sqrt{2}}\sin\left[\int_0^T\Omega(t)dt\right].
\end{equation*}
But from Eq. (\ref{limit_W}) we have
\begin{equation*}
\int_0^T\Omega(t)dt=\int_0^1\frac{1}{T\sqrt{2}}\frac{d\theta}{ds}Tds=\frac{1}{\sqrt{2}}\int_0^\pi d\theta=\frac{\pi}{\sqrt{2}},
\end{equation*}
and the fidelity limit is
\begin{equation}
\label{eta}
|c_2(T)|^2=\frac{1}{2}\sin^2\left(\frac{\pi}{\sqrt{2}}\right)\approx 0.3166,
\end{equation}
which agrees with the numerical value obtained from simulation. Note that this limit is the same for both the symmetric and nonsymmetric shortcuts, since in both cases angle $\theta$ changes by $\pi$.

The conclusion is that the TQD method requires several units of time ($\xi^{-1}$) in order to achieve acceptable levels of fidelity. The necessary duration is definitely lower than the time $T_a\approx30\xi^{-1}$ needed by the simple adiabatic method \cite{Unanyan01} to obtain comparable fidelity levels, see Fig. 5 in Ref. \cite{Paul16}, but it obviously cannot be reduced to the 1\% of $T_a$ reported in \cite{Paul16}. For $T=10\xi^{-1}=T_a/3$, the fidelities for the symmetric and nonsymmetric shortcuts are $0.9993$ and $0.9991$, respectively. In Figs. \ref{fig:delta_shortcut}, \ref{fig:omega_shortcut} we plot the controls for the two shortcuts, while in Figs. \ref{fig:end}, \ref{fig:middle} the corresponding evolution of populations $|c_1(t)|^2$ (solid line), $|c_2(t)|^2$ (dashed line), and $|c_3(t)|^2$ (dashed-dotted line).


\section{Optimal control}

\label{sec:OC}

In this section we follow an optimal control approach in order to obtain acceptable fidelity levels in shorter times. We use the freely available optimal control solver BOCOP \cite{bocop} to numerically solve a series of optimal control problems for the three-level system (\ref{SchrodingerC}), (\ref{Hc}),  with various durations $T$ and objective the maximization of $|c_2(T)|^2$, the final population of the triplet Bell state $|\psi^+_{\downarrow\uparrow}\rangle$.
Note that in the BOCOP software package, the continuous-time optimal control problem is approximated by a finite-dimensional optimization problem, using time discretization. The resultant nonlinear programming problem is subsequently solved using the nonlinear solver Ipopt. For the current problem we use a time discretization of 1000 points.

\begin{figure}[t]
 \centering
		\begin{tabular}{cc}
     	\subfigure[$\ $]{
	            \label{fig:D_0_T_2}
	            \includegraphics[width=.5\linewidth]{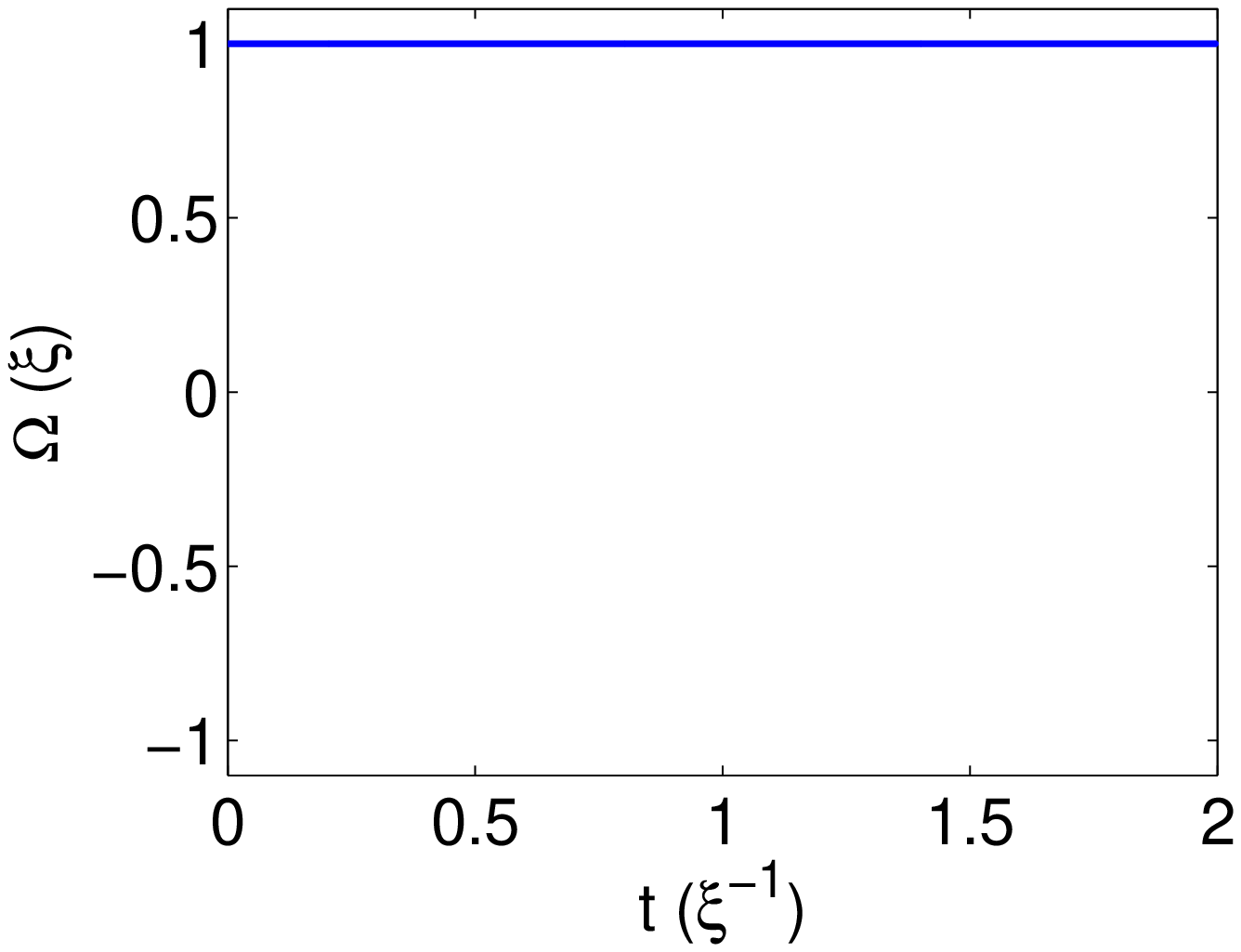}} &
       \subfigure[$\ $]{
	            \label{fig:popD_0_T_2}
	            \includegraphics[width=.5\linewidth]{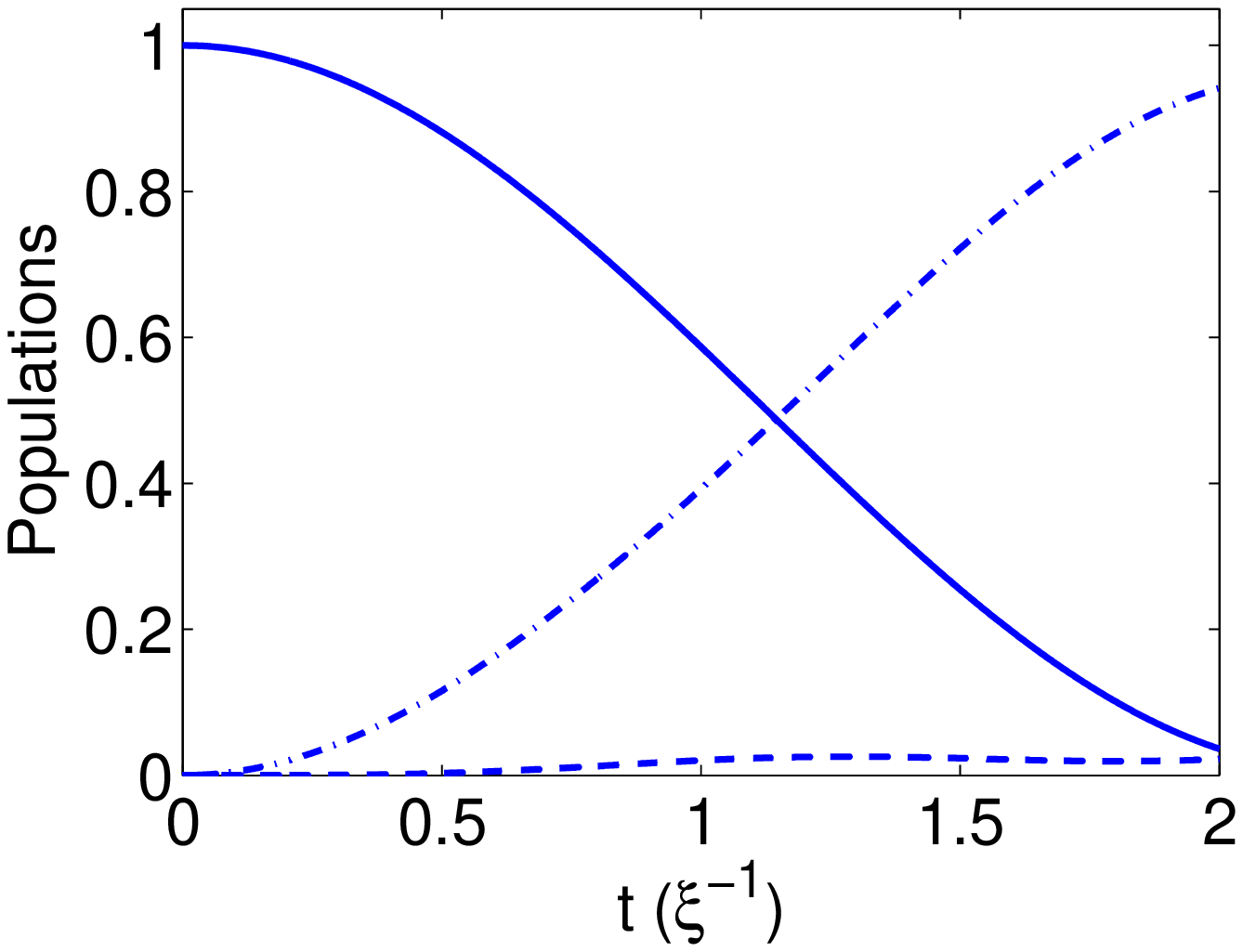}} \\
	    \subfigure[$\ $]{
	            \label{fig:D_0_T_25}
	            \includegraphics[width=.5\linewidth]{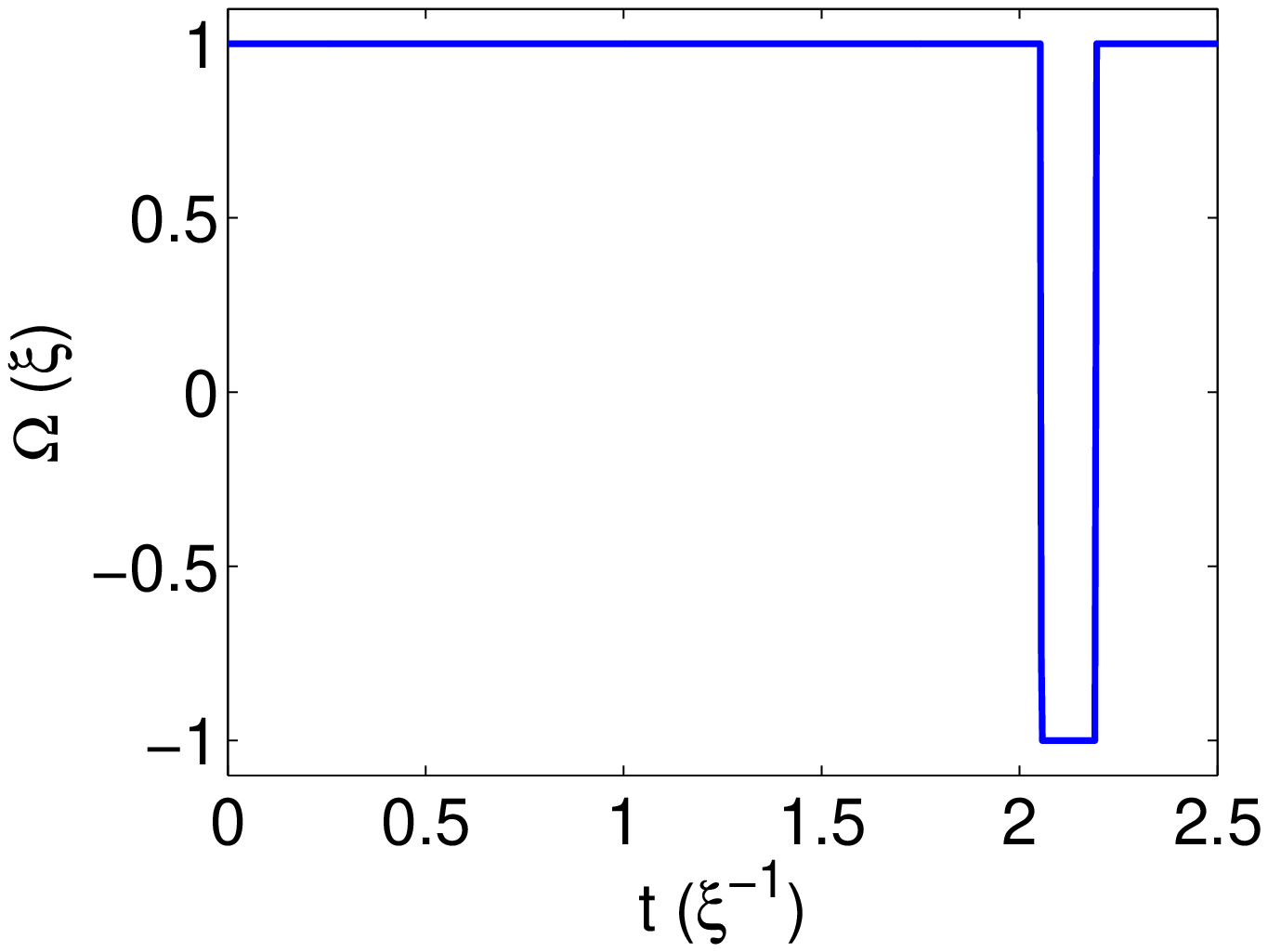}} &
       \subfigure[$\ $]{
	            \label{fig:popD_0_T_25}
	            \includegraphics[width=.5\linewidth]{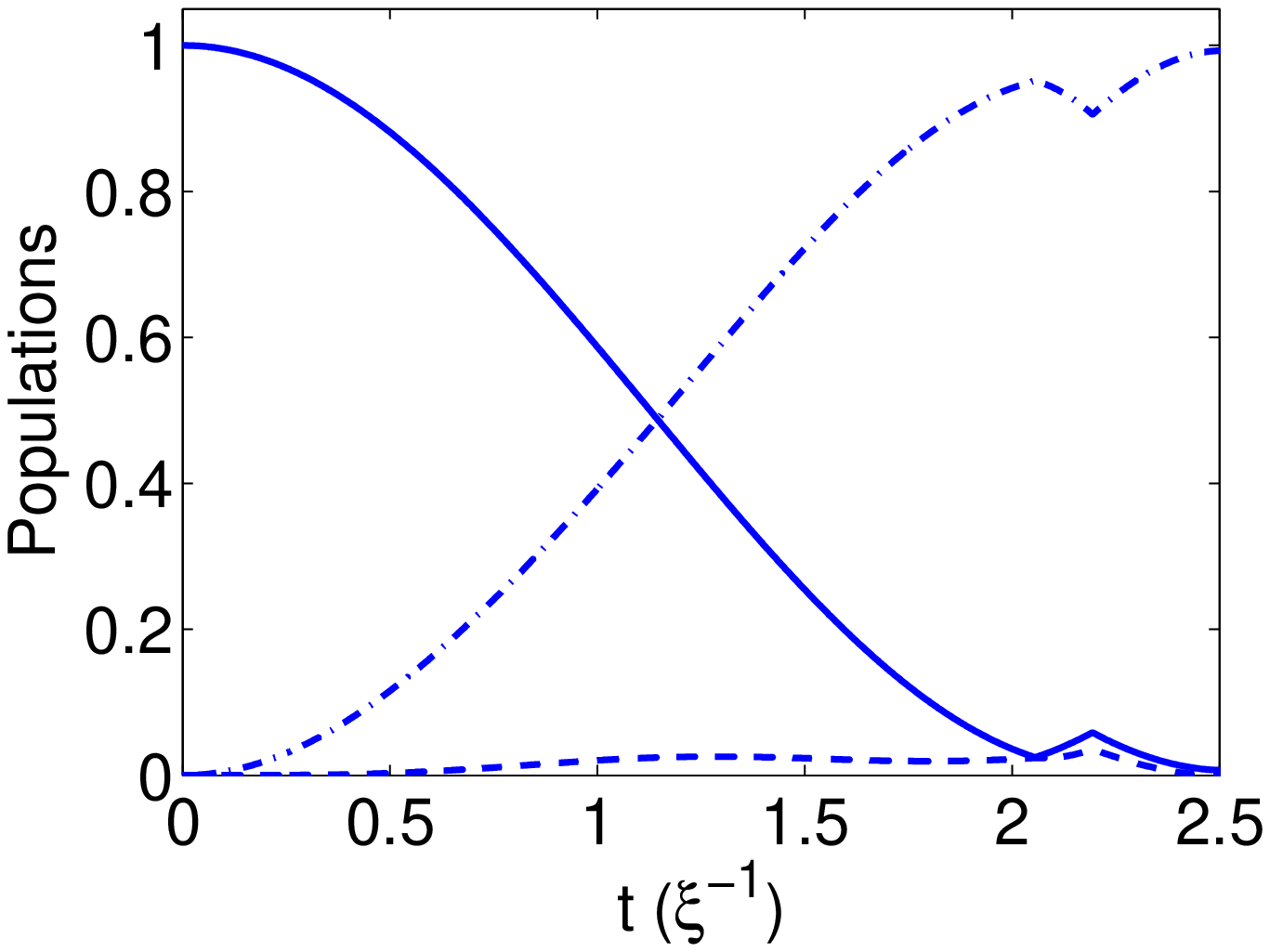}} \\
        \subfigure[$\ $]{
	            \label{fig:D_0_T_3}
	            \includegraphics[width=.5\linewidth]{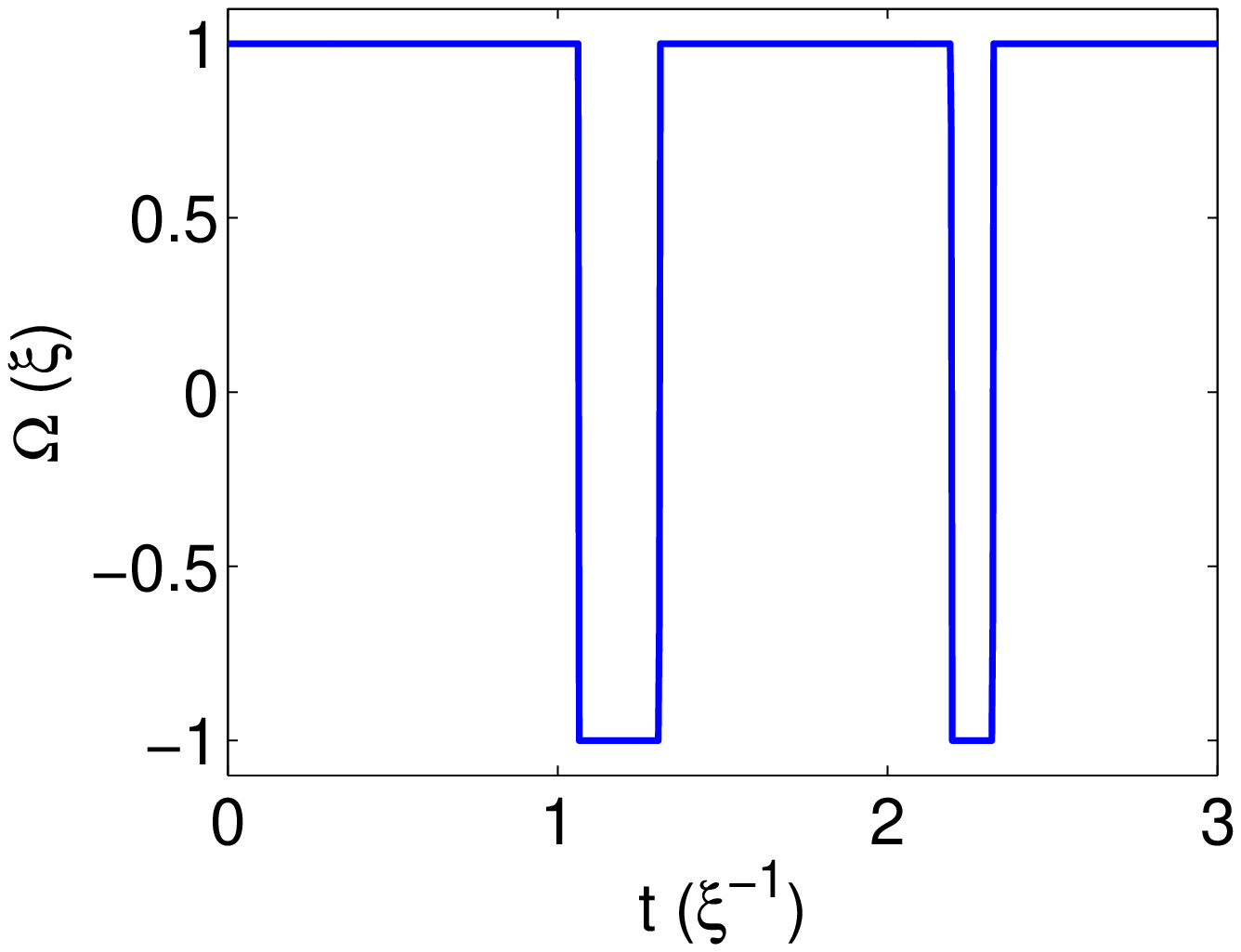}} &
       \subfigure[$\ $]{
	            \label{fig:popD_0_T_3}
	            \includegraphics[width=.5\linewidth]{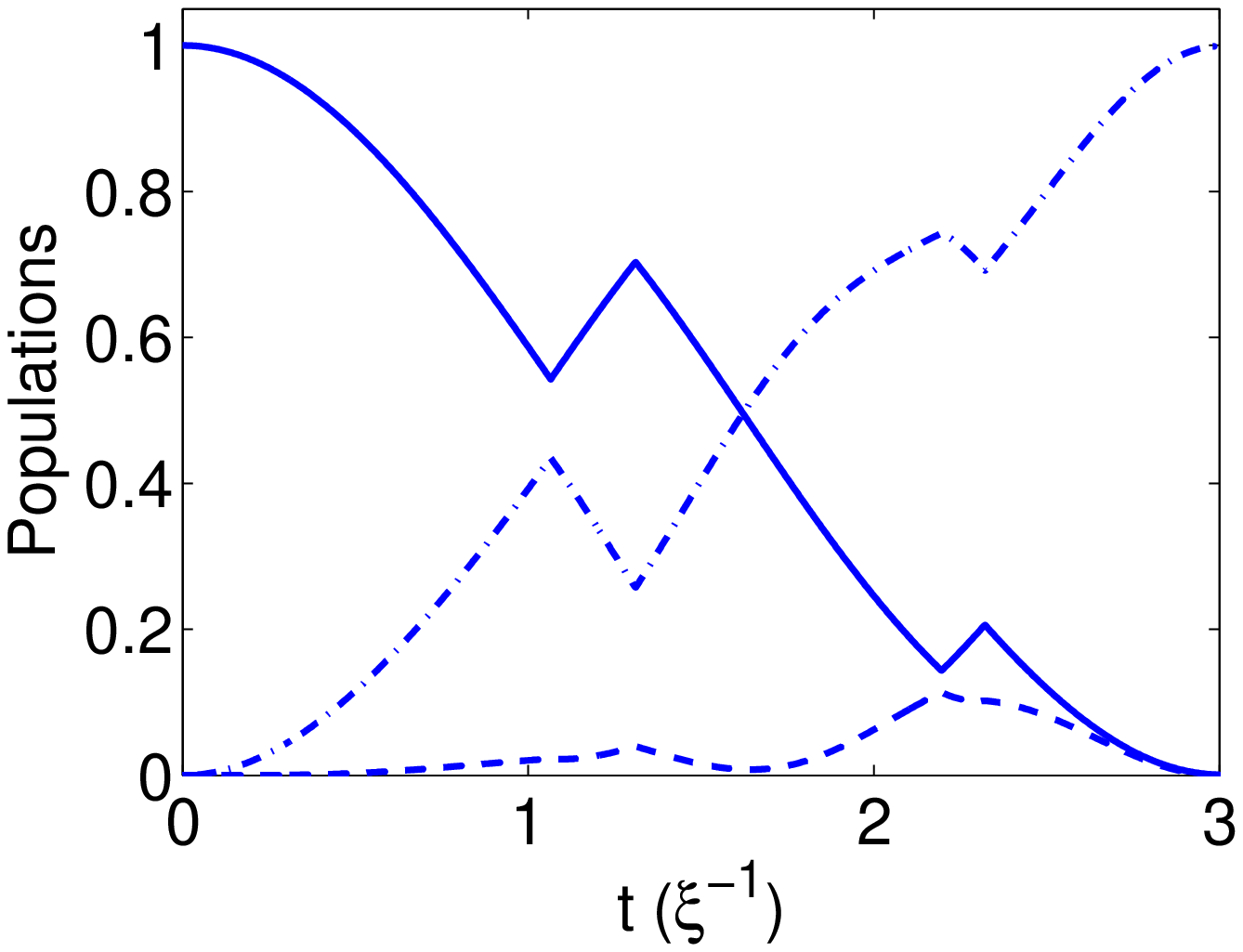}} \\
	    \subfigure[$\ $]{
	            \label{fig:D_0_T_36}
	            \includegraphics[width=.5\linewidth]{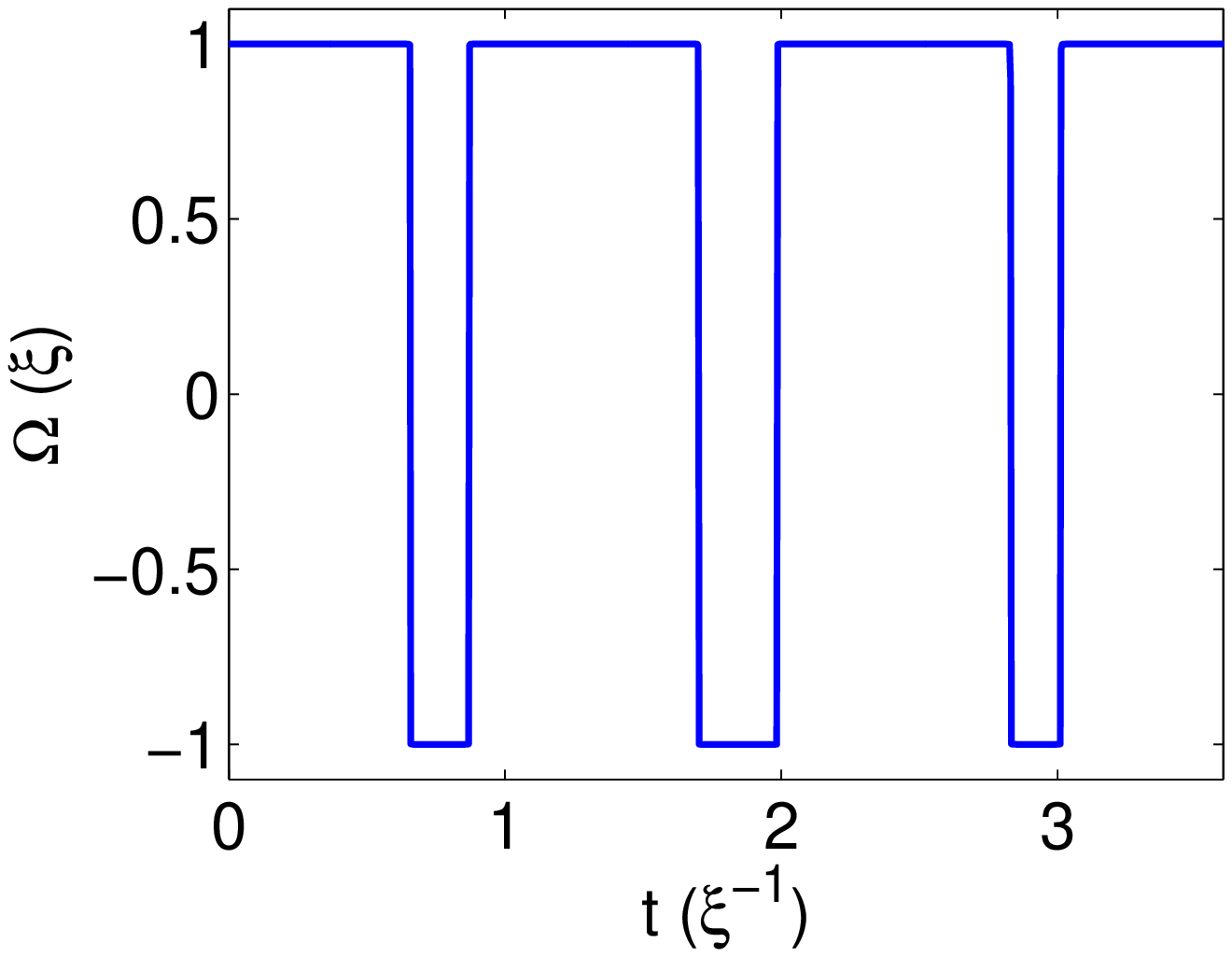}} &
       \subfigure[$\ $]{
	            \label{fig:popD_0_T_36}
	            \includegraphics[width=.5\linewidth]{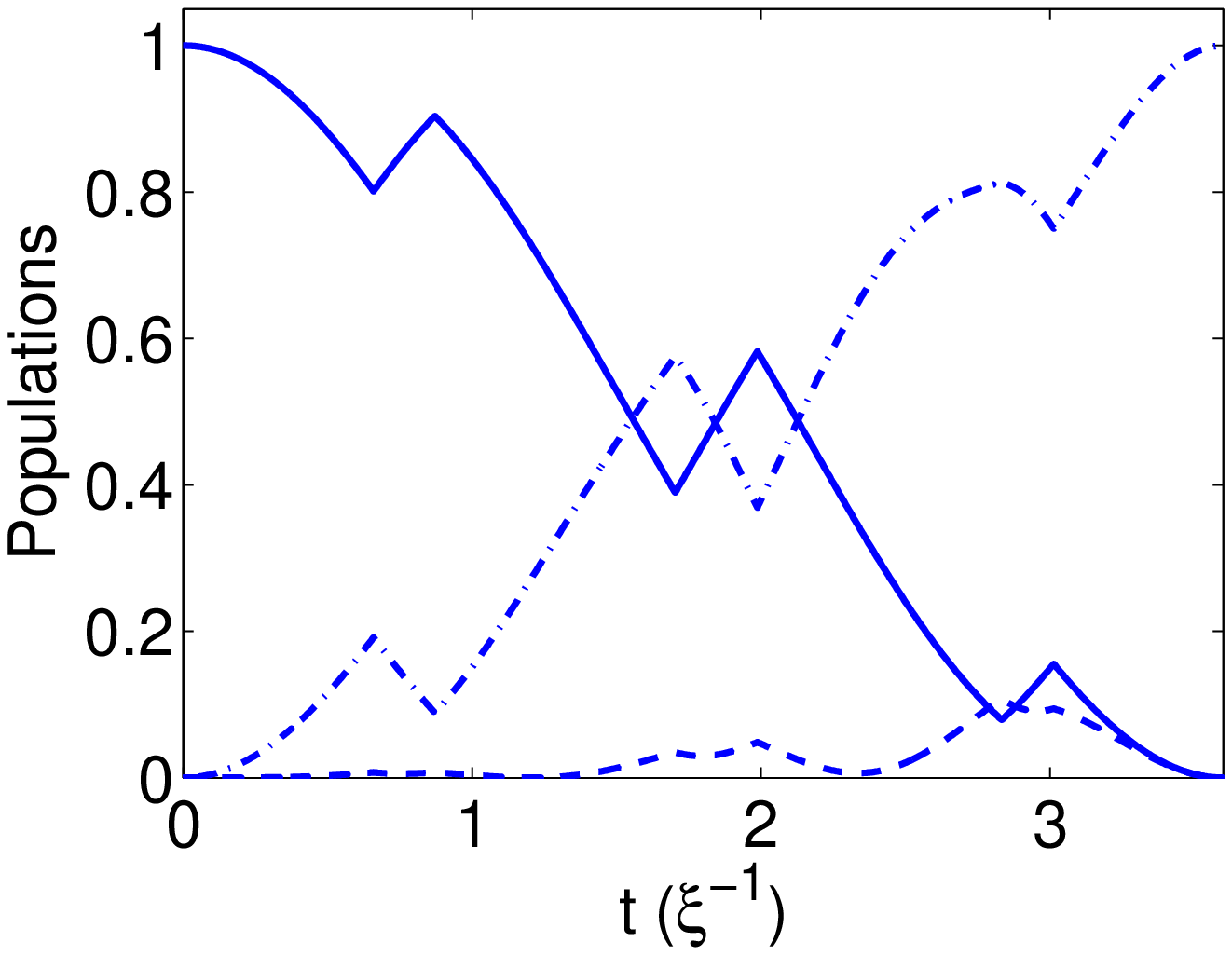}}
		\end{tabular}
\caption{(Color online) Optimal Rabi frequency $\Omega(t)$ for zero detuning $\Delta=0$ and various durations: (a) $T=2\xi^{-1}$, (c) $T=2.5\xi^{-1}$, (e) $T=3\xi^{-1}$, (g) $T=3.6\xi^{-1}$. The corresponding evolution of populations is displayed in (b,d,f,h). The fidelities for the depicted cases are 0.9416, 0.9928, 0.9990, 1.}
\label{fig:D_0}
\end{figure}

\begin{figure}[t]
 \centering
		\begin{tabular}{cc}
     	\subfigure[$\ $]{
	            \label{fig:fid_D}
	            \includegraphics[width=.5\linewidth]{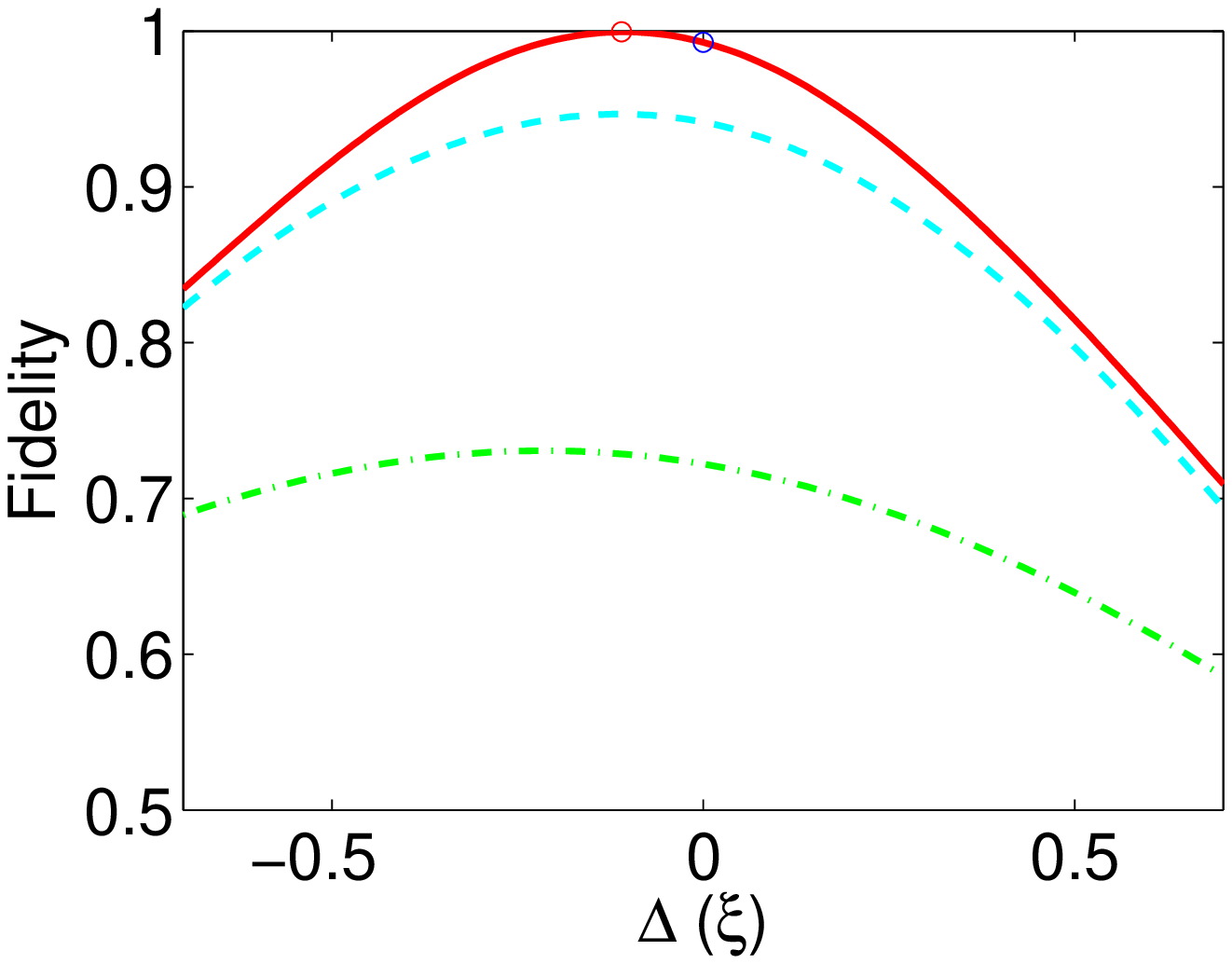}} &
       \subfigure[$\ $]{
	            \label{fig:fid_T}
	            \includegraphics[width=.5\linewidth]{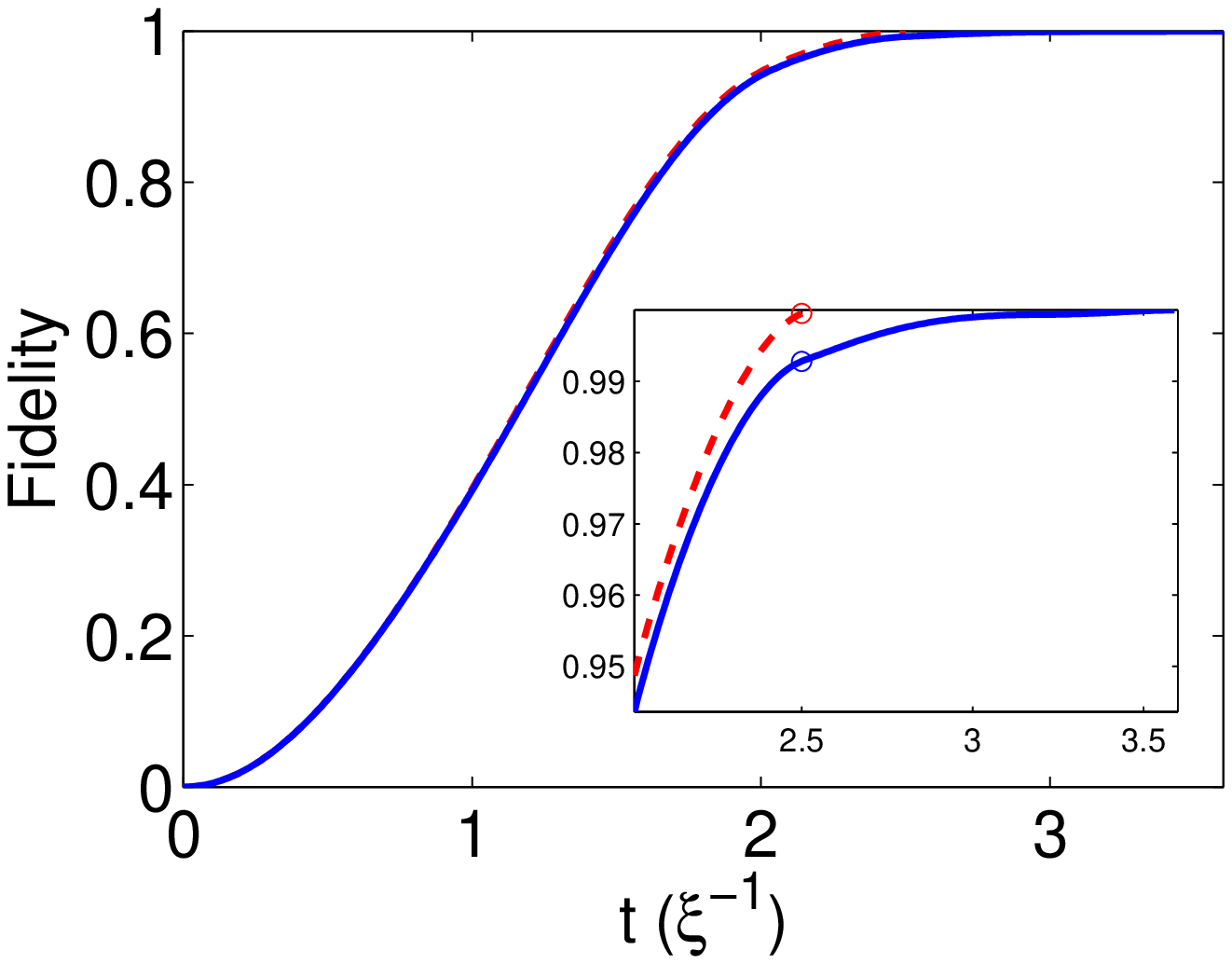}} \\
	    \subfigure[$\ $]{
	            \label{fig:D11_T_25}
	            \includegraphics[width=.5\linewidth]{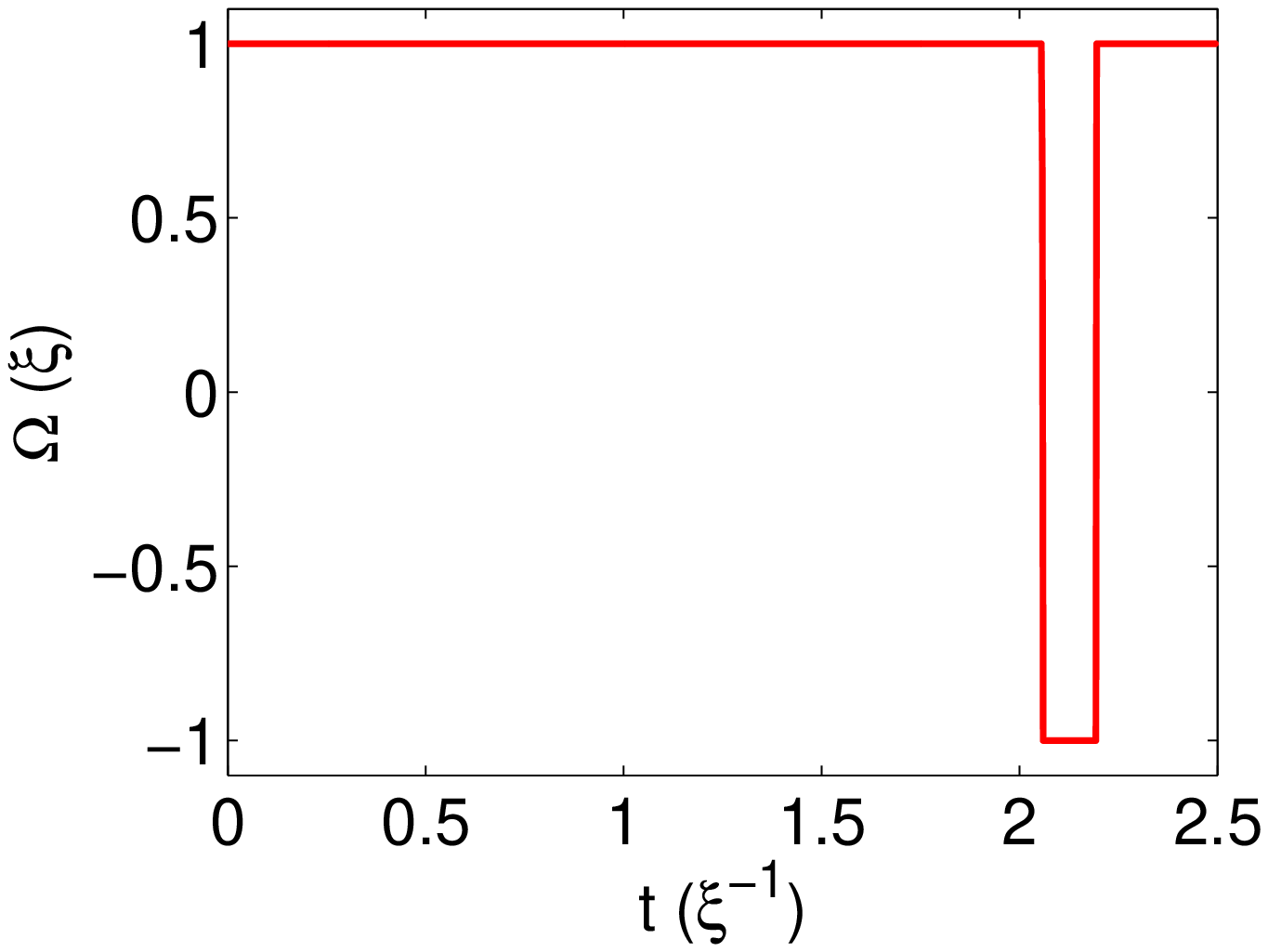}} &
       \subfigure[$\ $]{
	            \label{fig:popD11_T_25}
	            \includegraphics[width=.5\linewidth]{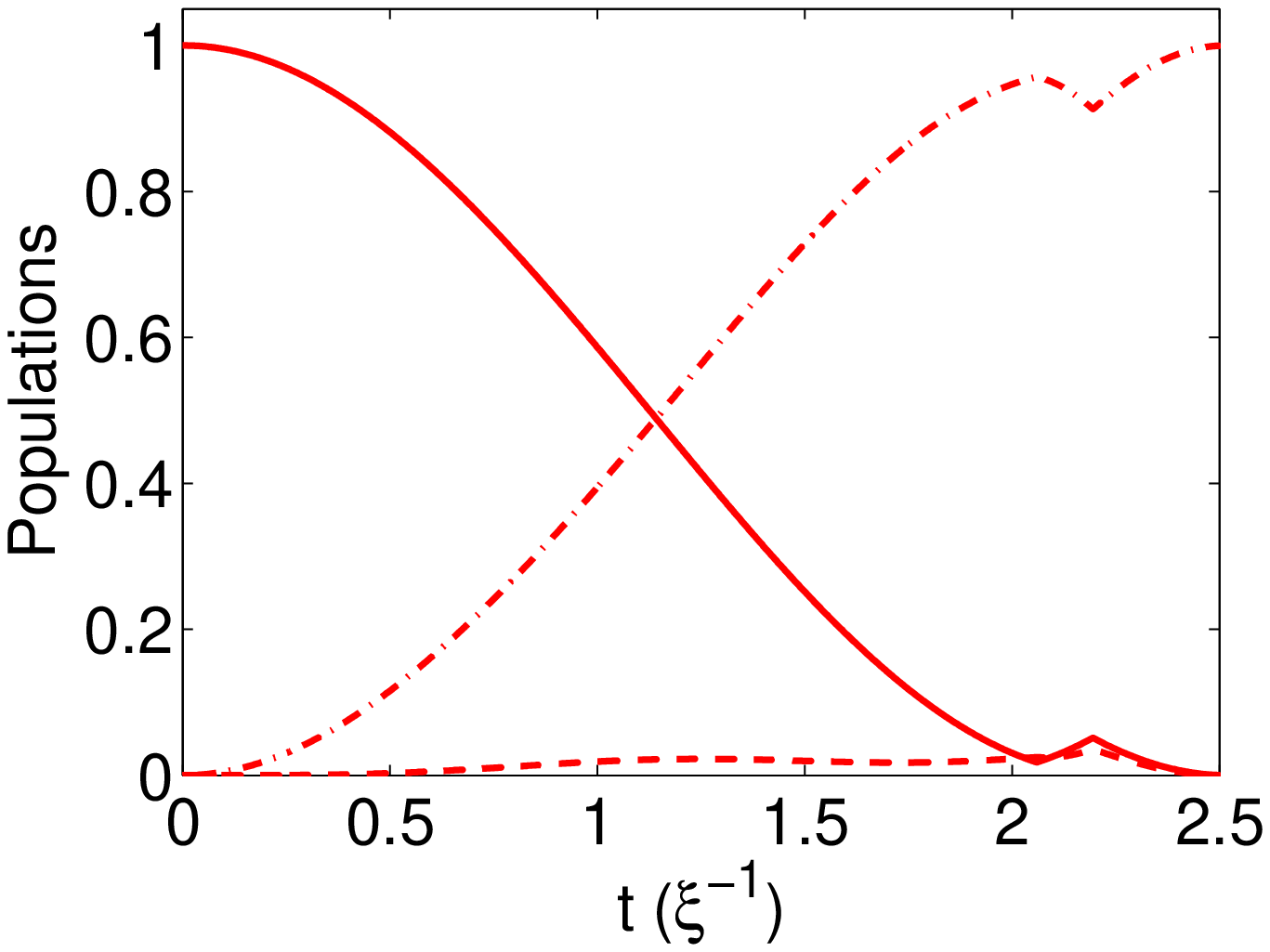}}
		\end{tabular}
\caption{(Color online) (a) Fidelity as a function of detuning $\Delta$, when $\Delta$ is kept constant and the Rabi frequency is optimized, for various durations $T=2.5\xi^{-1}$ (red solid line), $T=2\xi^{-1}$ (cyan dashed line), $T=1.5\xi^{-1}$ (green dashed-dotted line). The blue circle corresponds to the case with $\Delta=0$ depicted in Figs. \ref{fig:D_0_T_25}, \ref{fig:popD_0_T_25}, while the red circle correspond to the case depicted in Figs. \ref{fig:D11_T_25}, \ref{fig:popD11_T_25}. (b) Fidelity as a function of duration for optimized Rabi frequency and $\Delta=0$ (blue solid line), $\Delta=-0.11\xi$ (red dashed line). The inset shows that the case with negative detuning achieves faster the maximum efficiency. (c) Optimal Rabi frequency for the optimal constant detuning $\Delta=-0.11\xi$ when $T=2.5\xi^{-1}$. (d) Corresponding evolution of populations. The final fidelity is $0.9995$.}
\label{fig:D11}
\end{figure}

We initially fix the detuning to the constant value $\Delta=0$ and optimize the Rabi frequency $\Omega(t)$. Throughout this section we consider the bounds
\begin{equation}
\label{omega_bound}
-\xi\leq\Omega(t)\leq\xi.
\end{equation}
In Fig. \ref{fig:D_0} we plot the optimal $\Omega(t)$ and the corresponding time evolution of populations for various values of the duration $T$. Observe that the optimal Rabi frequency has the bang-bang form, where the signal alternates between the boundary values. For short durations the optimal control is a simple bang pulse, Fig. \ref{fig:D_0_T_2}, while for larger time intervals more bangs are introduced in order to further increase the fidelity, Figs. \ref{fig:D_0_T_25}, \ref{fig:D_0_T_3}, \ref{fig:D_0_T_36}. The fidelity as a function of duration $T$ is displayed in Fig. \ref{fig:fid_T} (blue solid line). Observe that here, contrary to the TQD case, the fidelity vanishes as $T$ approaches zero. This happens because now the control is bounded, see Eq. (\ref{omega_bound}), while in the TQD case becomes a delta pulse for small $T$.

The more complicated switching structures shown in Fig. \ref{fig:D_0}, which are necessary in order to increase the fidelity of the final state, might be difficult to accurately implement experimentally. In order to reach the same fidelity levels with more tractable controls, we follow alternative approaches. One simple idea is to try \emph{constant} values of detuning different than $\Delta=0$ used before, and optimize $\Omega(t)$ for them. In Fig. \ref{fig:fid_D} we plot the fidelity as a function of detuning $\Delta$, when $\Delta$ is kept constant in time and the Rabi frequency is optimized, for various durations $T=2.5\xi^{-1}$ (red solid line), $T=2\xi^{-1}$ (cyan dashed line), $T=1.5\xi^{-1}$ (green dashed-dotted line). Observe that in all the depicted cases the best efficiency is obtained for $\Delta<0$. This can be intuitively understood by inspection of Eq. (\ref{Hc}), where obviously a small negative $\Delta$ increases the detuning $4\xi-\Delta$ of the undesirable transfer $|\psi^+_{\downarrow\uparrow}\rangle\rightarrow|\psi_{\upuparrows}\rangle$  while is affecting less the desired transfer $|\psi_{\downdownarrows}\rangle\rightarrow|\psi^+_{\downarrow\uparrow}\rangle$. In Fig. \ref{fig:fid_T} we plot the fidelity as a function of duration $T$ for fixed detuning $\Delta=-0.11\xi$ (red dashed line). The inset demonstrates that now a very good efficiency is obtained faster, compared to the case where $\Delta=0$ (blue solid line). In Fig. \ref{fig:D11_T_25} we depict the optimal $\Omega(t)$ for fixed $\Delta=-0.11\xi$ and duration $T=2.5\xi^{-1}$, the case highlighted with a red circle in Figs. \ref{fig:fid_D}, \ref{fig:fid_T}, while Fig. \ref{fig:popD11_T_25} displays the corresponding evolution of populations. Observe that the optimal pulse sequence contains only one negative bang, in analogy with the same duration case shown in Fig. \ref{fig:D_0_T_25} for $\Delta=0$, but the obtained fidelity is much larger, 0.9995 compared to 0.9928. It is actually comparable with the fidelity obtained with the more complicated pulse sequence shown in Fig. \ref{fig:D_0_T_36} for a longer duration $T=3.6\xi^{-1}$.

\begin{figure}[t]
 \centering
		\begin{tabular}{cc}
     	\subfigure[$\ $]{
	            \label{fig:p_200}
	            \includegraphics[width=.5\linewidth]{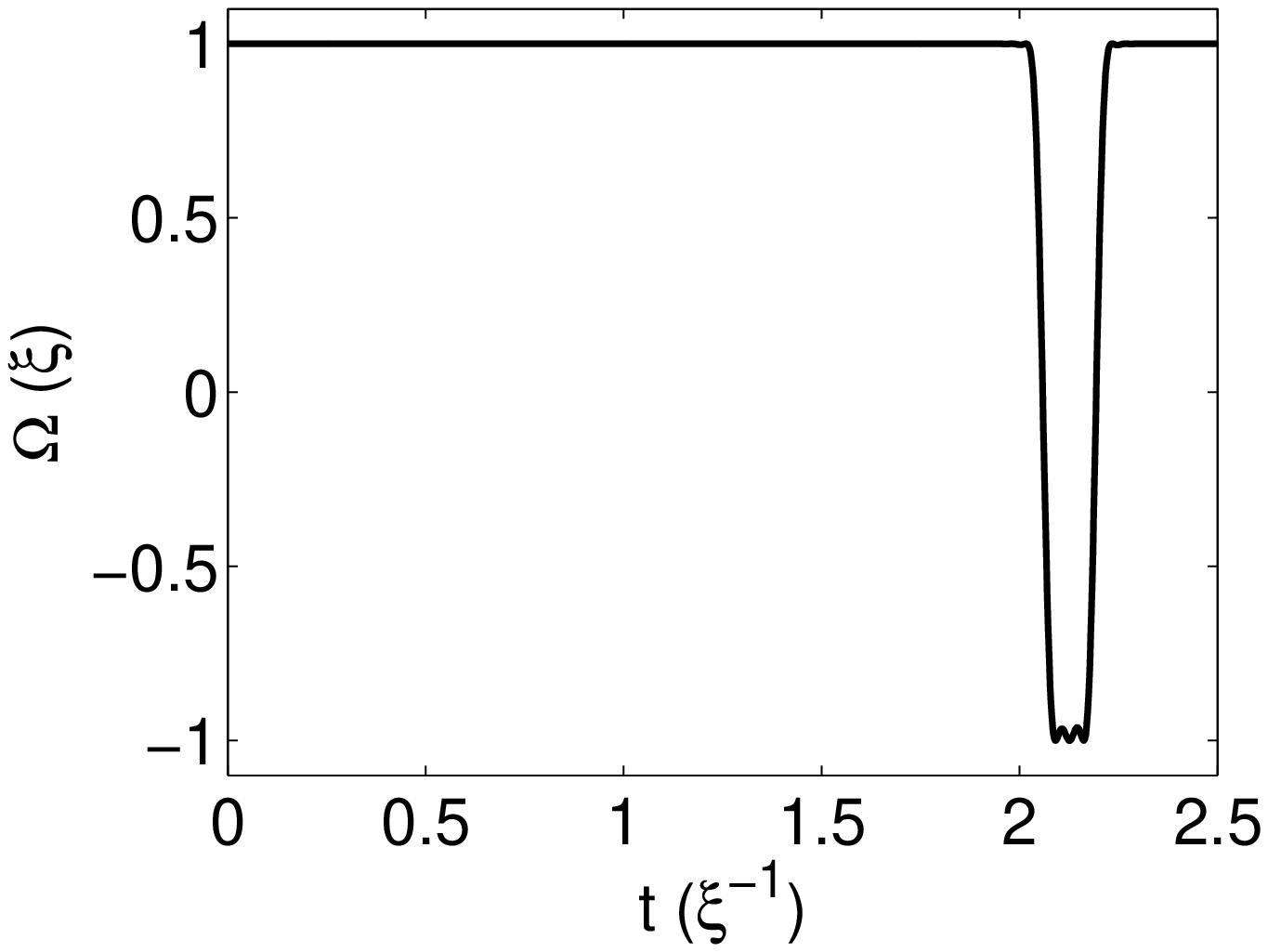}} &
       \subfigure[$\ $]{
	            \label{fig:popp_200}
	            \includegraphics[width=.5\linewidth]{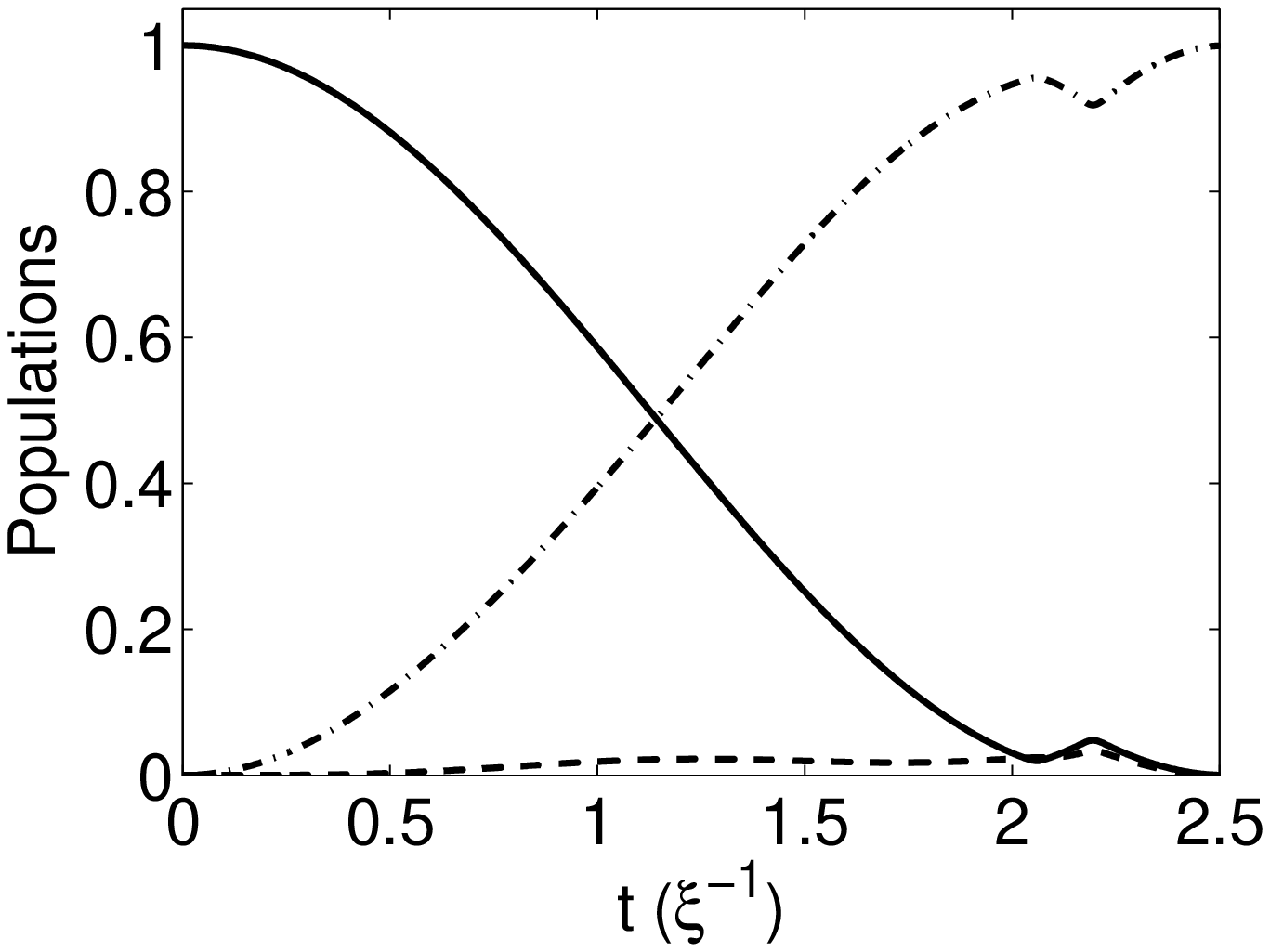}} \\
	    \subfigure[$\ $]{
	            \label{fig:fid_harmonics}
	            \includegraphics[width=.5\linewidth]{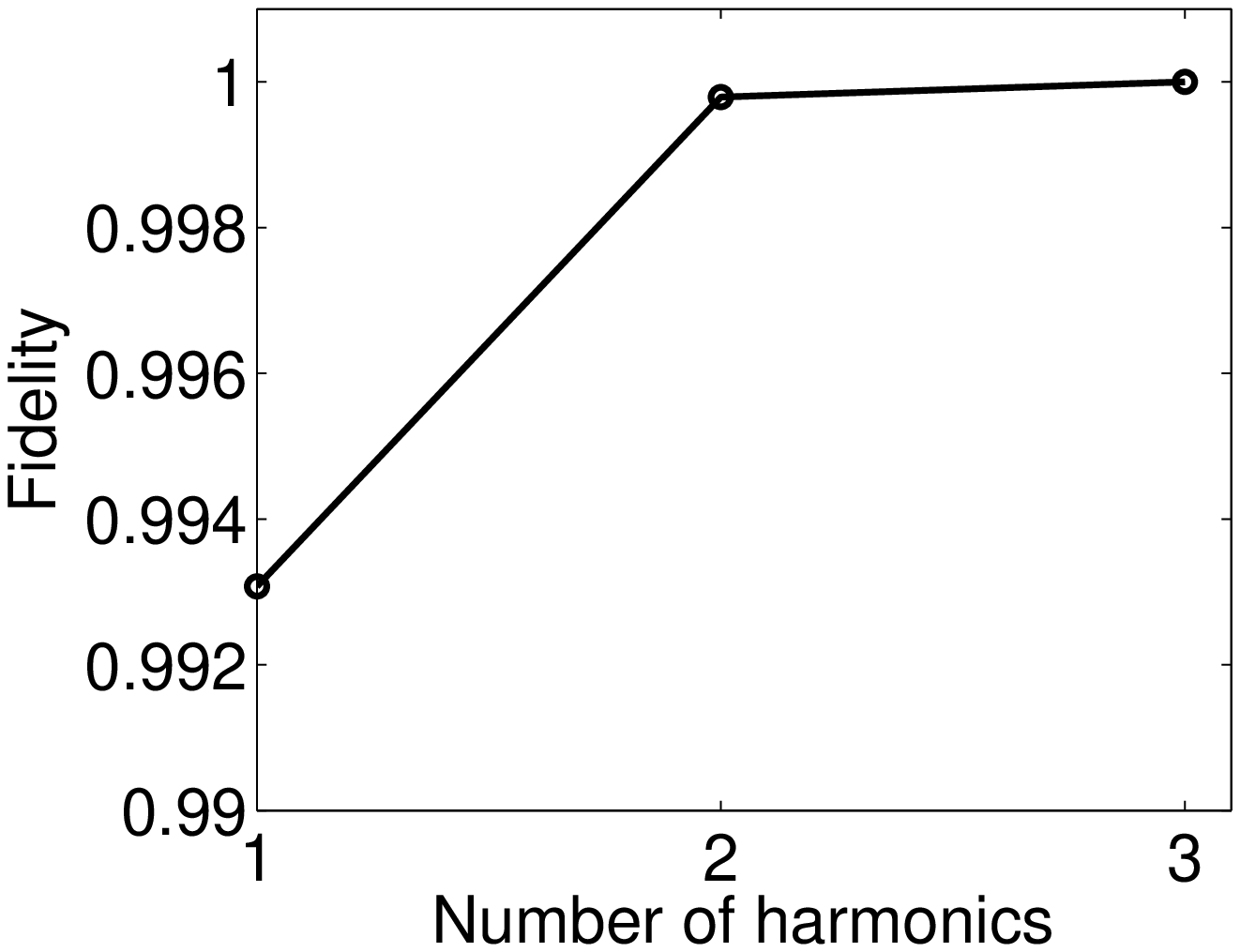}} &
       \subfigure[$\ $]{
	            \label{fig:D_p_3}
	            \includegraphics[width=.5\linewidth]{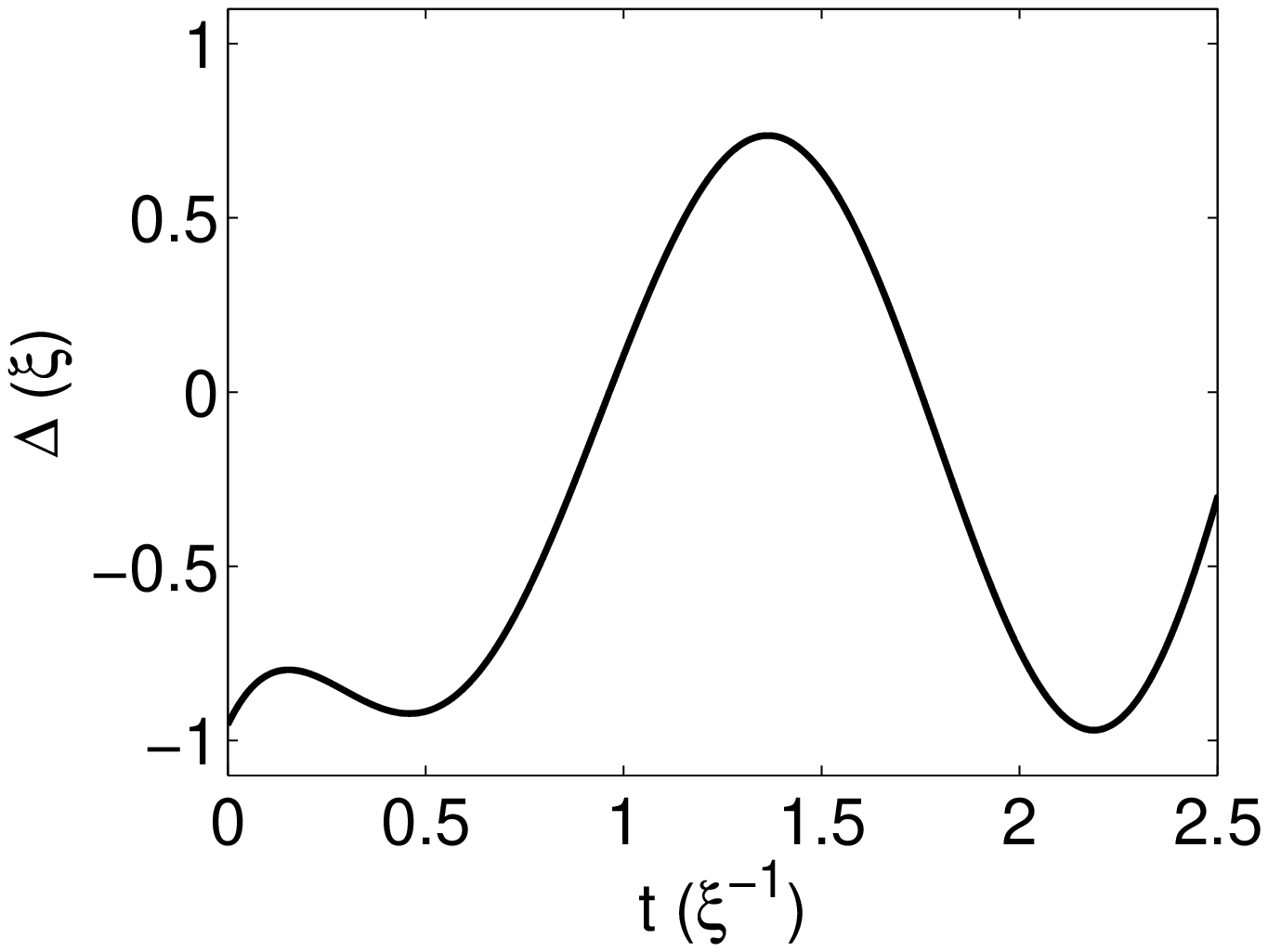}} \\
       \subfigure[$\ $]{
	            \label{fig:W_p_3}
	            \includegraphics[width=.5\linewidth]{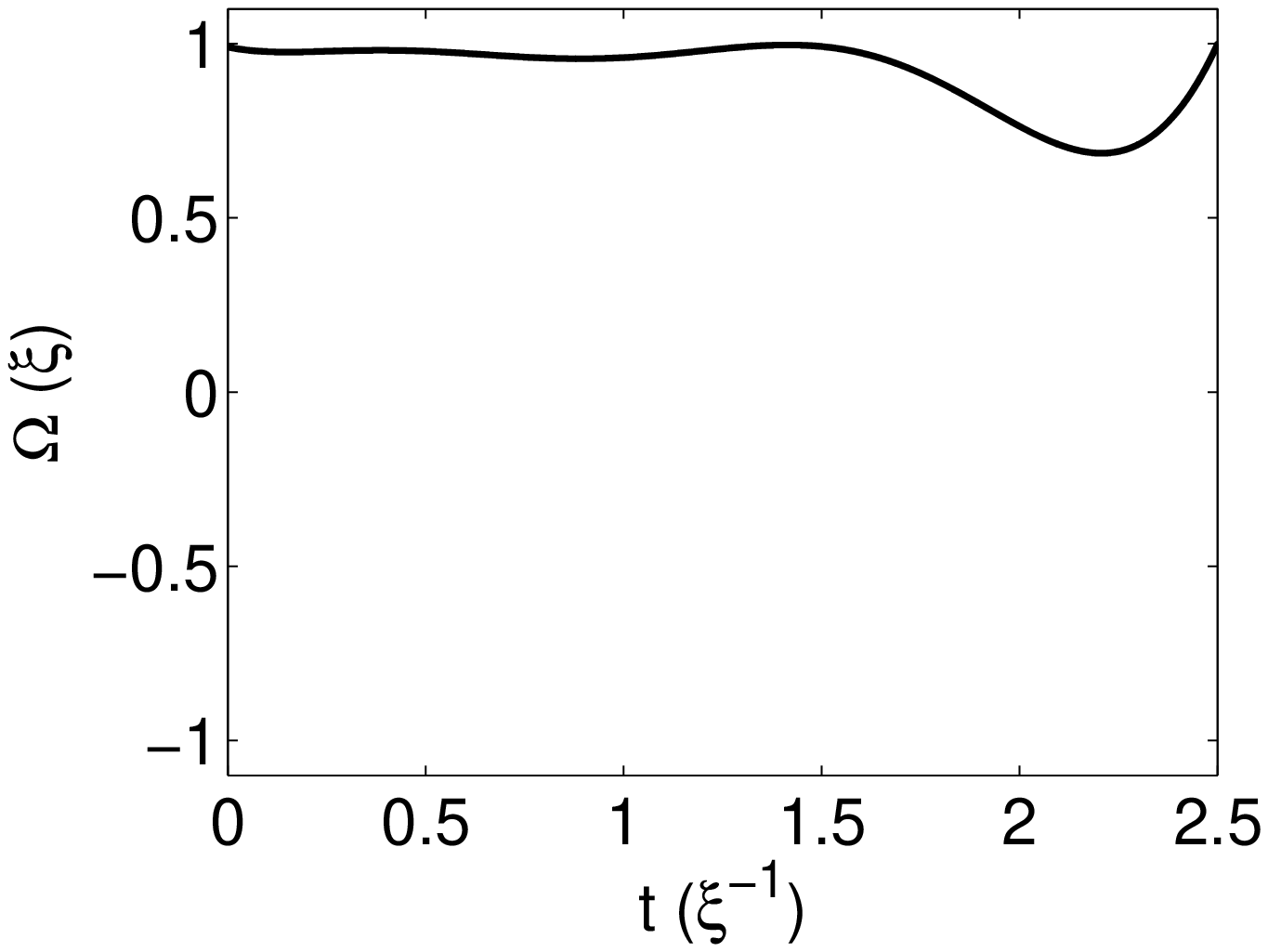}} &
       \subfigure[$\ $]{
	            \label{fig:popp_3}
	            \includegraphics[width=.5\linewidth]{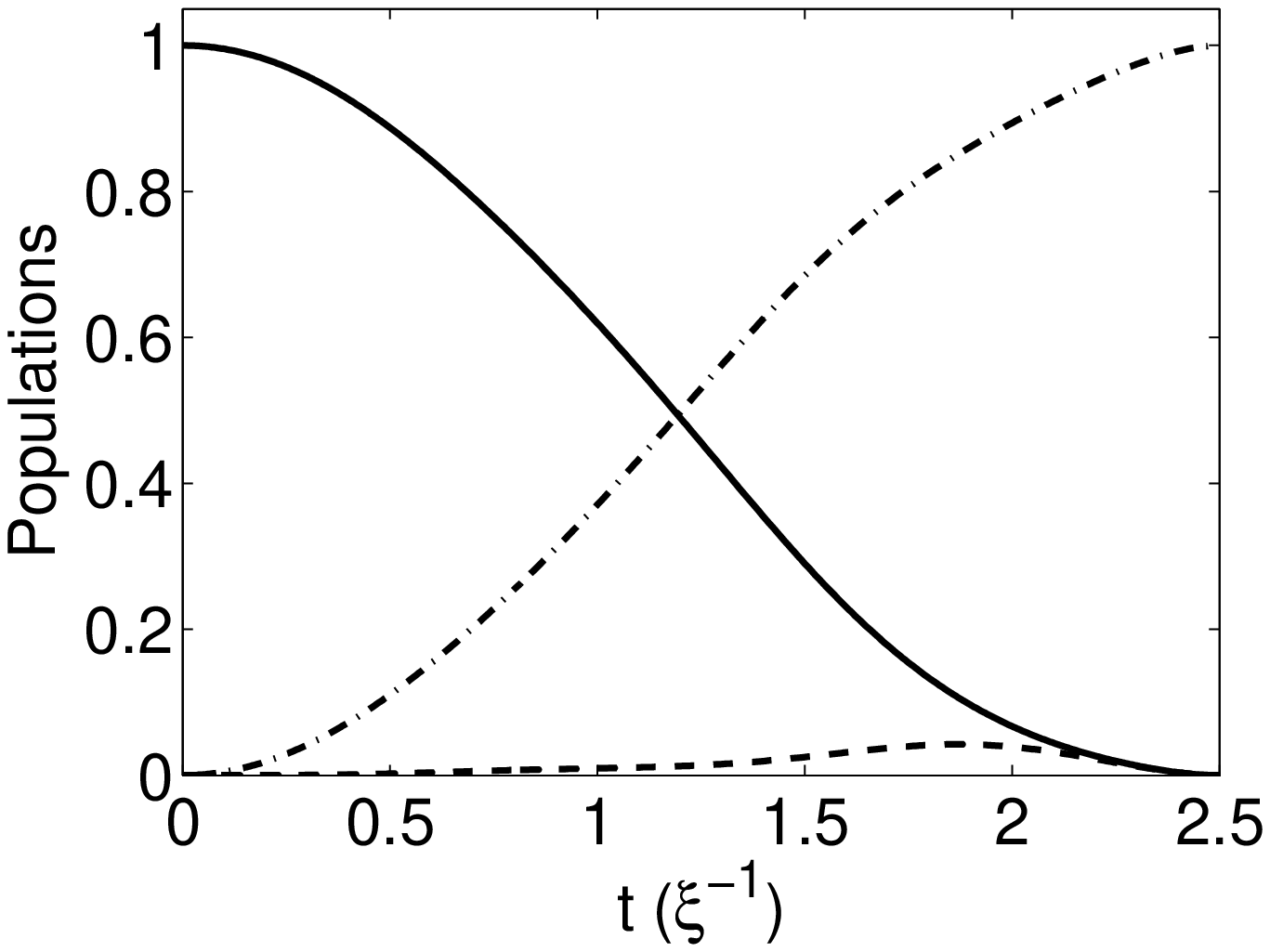}}
		\end{tabular}
\caption{(Color online) (a) Optimal Rabi frequency $\Omega(t)$ of the trigonometric form with $p=200$ harmonics, for fixed $\Delta=-0.11\xi$ and duration $T=2.5\xi^{-1}$. Observe that it approaches the optimal bang-bang form of Fig. \ref{fig:D11_T_25}. (b) Corresponding evolution of populations, similar to Fig. \ref{fig:popD11_T_25}. (c) Fidelity as a function of the number of harmonics when both $\Delta(t), \Omega(t)$ have the trigonometric form and are optimized for duration $T=2.5\xi^{-1}$. (d) Optimal trigonometric $\Delta(t)$ with $p=3$ harmonics and $T=2.5\xi^{-1}$. (e) Optimal trigonometric $\Omega(t)$ with $p=3$ harmonics and $T=2.5\xi^{-1}$. (f) Corresponding evolution of populations.}
\label{fig:sinus}
\end{figure}

We next move to find smooth optimal controls, probably more relevant for a possible experimental implementation, which can achieve comparable fidelity within the same time interval $T=2.5\xi^{-1}$. For this purpose we exploit a BOCOP feature which allows to seek optimal controls in a trigonometric series form, namely
\begin{subequations}
\label{sinus}
\begin{eqnarray}
\Omega(t)&=&a_0+\sum_{k=1}^p(a_{2k-1}\cos{kt}+a_{2k}\sin{kt}),\label{omega_sinus}\\
\Delta(t)&=&b_0+\sum_{k=1}^p(b_{2k-1}\cos{kt}+b_{2k}\sin{kt}).\label{delta_sinus}
\end{eqnarray}
\end{subequations}
In order to test the method, we first fix the detuning to the previously obtained optimal constant value $\Delta=-0.11\xi$ and optimize $\Omega(t)$ under constraints (\ref{omega_bound}) and (\ref{omega_sinus}) for duration $T=2.5\xi^{-1}$. In Fig. \ref{fig:p_200} we plot the optimal control obtained using $p=200$ harmonics in Eq. (\ref{omega_sinus}), while Fig. \ref{fig:popp_200} shows the corresponding evolution of populations. Observe the similarity to the case with a single negative bang displayed in Figs. \ref{fig:D11_T_25} and \ref{fig:popD11_T_25}, while the obtained fidelity is about the same, 0.9995. The necessary number of harmonics to reach this fidelity is quite large. In order to overcome this problem, we allow the time variation of the detuning and seek optimal $\Omega(t), \Delta(t)$ in the form (\ref{omega_sinus}), (\ref{delta_sinus}), under the constraint (\ref{omega_bound}) and a similar one for $\Delta(t)$
\begin{equation}
\label{delta_bound}
-\xi\leq\Delta(t)\leq\xi.
\end{equation}
Fig. \ref{fig:fid_harmonics} shows the fidelity achieved with this approach as a function of the number of harmonics used in the series (\ref{omega_sinus}) and (\ref{delta_sinus}). Observe that, when both $\Omega(t), \Delta(t)$ are optimized, a very good efficiency is already obtained with only two harmonics. In Figs. \ref{fig:W_p_3}, \ref{fig:D_p_3} we display the optimal $\Omega(t), \Delta(t)$ when $p=3$, and in Fig. \ref{fig:popp_3} the corresponding evolution of populations. A nearly perfect fidelity is obtained with these smooth controls. The optimal coefficients $a_k, b_k$ for the series (\ref{omega_sinus}), (\ref{delta_sinus}) with $p=3$ are given in Table \ref{tab:coefficients}.


\begin{table}[t!]
\caption{\label{tab:coefficients} Optimal coefficients for the trigonometric series (\ref{omega_sinus}), (\ref{delta_sinus}) when $p=3$ and the duration is set to $T=2.5\xi^{-1}$.}
\begin{ruledtabular}
\begin{tabular}{cc}
\textrm{$a_k$}&
\textrm{$b_k$}\\
\colrule
4.88177 & -8.67328\\
-3.02932 & 0.800026\\
-5.61925 & 14.4413\\
-1.64576 & 8.33812\\
2.79904 & -1.43694\\
0.784017 & -1.41904\\
-0.0724018 & -3.07217
\end{tabular}
\end{ruledtabular}
\end{table}

\section{Conclusion}

\label{sec:conclusion}

In this article, we studied the problem of efficient generation of the triplet Bell state in a system of two spins with Ising interaction. We started with the TQD method and
showed that its fidelity cannot approach unity in arbitrarily short times, as it usually happens for shortcut to adiabaticity methods. Then, we used numerical optimal control to obtain bang-bang pulse sequences and smooth controls which can create sufficient amount of this state in the shortest possible time. The current results are not restricted only to spin systems, but is also expected to find applications in other physical systems which can be modeled as interacting spins, as, for example, coupled quantum dots.

\begin{acknowledgments}
The research is implemented through the Operational Program ``Human Resources Development, Education and Lifelong Learning'' and is co-financed by the European Union (European Social Fund) and Greek national funds.
\end{acknowledgments}

\end{document}